\documentclass[prd,aps,twocolumn,nofootinbib,floatfix,longbibliography]{revtex4-1}
\usepackage{graphicx}
\usepackage{bm}
\usepackage{mathtools}
\usepackage{hyperref}
\usepackage{mathrsfs}
\newcommand\unit[1]{{\rm #1}}
\newcommand\E[1]{{\left\langle #1 \right \rangle}}

\usepackage{color}

\newcommand\nSimStart{36}
\newcommand\nSimPlaced{448}
\newcommand\tStartDays{1/8}
\newcommand\tEndDays{7.66}
\newcommand\nTimePoints{191}

\newcommand{\aj}{Astronomical  Journal}
\newcommand{\apjl}{Astrophysical Journal Letters}
\newcommand{\apjs}{Astroph. J. S.}

\newcommand{\pasj}{PASJ}
\newcommand{\jcap}{JCAP}
\newcommand{\mnras}{Monthly Notices of the Royal Astronomical Society}
\newcommand{\aap}{Astronomy and Astrophysics}

\def\RIT{Center for Computational Relativity and Gravitation, Rochester Institute of Technology, Rochester, New York
  14623, USA}
\def\XCP{Computational Physics Division, Los Alamos National Laboratory, Los Alamos, NM, 87545, USA}
\def\CTA{Center for Theoretical Astrophysics, Los Alamos National Laboratory, Los Alamos, NM, 87545, USA}
\def\CCSS{Computer, Computational, and Statistical Sciences Division,  Los Alamos National Laboratory, Los Alamos, NM,
  87545, USA}
\def\UA{The University of Arizona, Tucson, AZ 85721, USA}
\def\NM{Department of Physics and Astronomy, The University of New Mexico, Albuquerque, NM 87131, USA}
\def\CIERA{Center for Interdisciplinary Exploration and Research in Astrophysics (CIERA), Northwestern University, Evanston, IL, 60201, USA}
\def\NW{Department of Physics and Astronomy, Northwestern University, Evanston, IL 60208, USA}

\begin{document}

\title{Interpolating detailed simulations of kilonovae: adaptive learning and parameter inference applications}

\author{M. Ristic}
\affiliation{\RIT}
\author{E. Champion}
\affiliation{\RIT}
\author{R. O'Shaughnessy}
\affiliation{\RIT}
\author{R. Wollaeger}
\affiliation{\CTA}
\affiliation{\CCSS}
\author{O. Korobkin}
\affiliation{\CTA}
\affiliation{\CCSS}
\author{E.~A. Chase}
\affiliation{\CTA}
\affiliation{\XCP}
\affiliation{\CIERA}
\affiliation{\NW}
\author{C.~L. Fryer}
\affiliation{\CTA}
\affiliation{\CCSS}
\affiliation{\UA}
\affiliation{\NM}
\author{A.~L. Hungerford}
\affiliation{\CTA}
\affiliation{\XCP}
\author{C.~J. Fontes}
\affiliation{\CTA}
\affiliation{\XCP}

\date{\today}

\begin{abstract}
Detailed radiative transfer simulations of kilonovae are difficult to apply directly to observations; they only
sparsely cover simulation parameters, such as the mass, velocity, morphology, and composition of the ejecta.    On the other hand,
semianalytic models for kilonovae can be evaluated continuously over model parameters, but neglect important %
physical details which are not incorporated in the simulations, thus introducing systematic bias.
Starting with a grid of 2D anisotropic simulations of kilonova light curves covering a wide range of ejecta properties, we apply adaptive-learning techniques to iteratively
choose new simulations and produce high-fidelity surrogate models for those simulations. These surrogate models allow for continuous evaluation across model parameters
while retaining the microphysical details about the ejecta.
Using a new code for multimessenger inference, we demonstrate how to use our interpolated models to infer
kilonova parameters.  %
Comparing to inferences using simplified analytic models, we recover different ejecta properties.
We discuss the implications of this analysis which is qualitatively consistent with similar previous work using
detailed ejecta opacity calculations and which illustrates systematic challenges for kilonova modeling.
An associated data and code release provides our interpolated light-curve models, interpolation implementation which can be
applied to reproduce our work or extend to new models, and our multimessenger parameter inference engine.
\end{abstract}

\maketitle

\section{Introduction}
As exemplified by GW170817, neutron star mergers are empirically known to produce a rich array of multimessenger
emission \cite{2017ApJ...848L..12A,LIGO-GW170817-bns}.  The presence of matter is most unambiguously indicated by electromagnetic
emission from nuclear matter ejected during the merger itself, which produces distinctive ``kilonova''
emission \cite{1974ApJ...192L.145L,1998ApJ...507L..59L,2019LRR....23....1M,2020GReGr..52..108B} via radioactive heating of this expanding
 material.  
Kilonova observations can provide insight into uncertain nuclear physics \cite{2020arXiv201011182B,2021ApJ...906...94Z,2020arXiv200604322V,2019AnPhy.41167992H,2020GReGr..52..109C} and help constrain the expansion rate of
the universe  
\cite{2020arXiv201101211C,2020NatCo..11.4129C,2020PhRvR...2b2006C,2020ApJ...892L..16D},
particularly in conjunction with gravitational wave observations 
\cite{LIGO-GW170817-mma,LIGO-GW170817-H0,LIGO-GW170817-EOS,LIGO-GW170817-EOSrank,2020PhRvL.125n1103B,2019LRR....23....1M,2019NatAs...3..940H,2020Sci...370.1450D}.

In principle,  kilonova observations encode the amount and properties of the ejected material in their complex
multi-wavelength light curves (and spectra)  \cite{2019LRR....23....1M,2018MNRAS.480.3871C,2017ApJ...851L..21V}.  
For example, several studies of GW170817 attempted to infer the amount of material ejected
\cite{2021arXiv210101201B,gwastro-mergers-em-CoughlinGPKilonova-2020,2018MNRAS.480.3871C,2019MNRAS.489L..91C,2017ApJ...851L..21V,2017Natur.551...75S,tanvir17,2017ApJ...848L..21A,chornock17,2017ApJ...848L..17C}.
In practice, these observations have historically been interpreted with semianalytic models, as they can be evaluated quickly and
continuously over the parameters which characterize potential merger ejecta.   
However, it is well known that these semianalytic models contain oversimplified physics of already simplified anisotropic
radiative transfer calculations  \cite{2018MNRAS.478.3298W,2020ApJ...899...24E,kilonova-lanl-WollaegerNewGrid2020} that neglect
detailed anisotropy, radiative transfer, opacity, sophisticated nuclear reaction networks, and composition differences.

To circumvent these biases, some groups have attempted to construct surrogate kilonova light-curve models, calibrated to
detailed radiative transfer simulations
\cite{gwastro-mergers-em-CoughlinGPKilonova-2020,2018MNRAS.480.3871C,RisticThesis}.
For example, Coughlin et al. \cite{2018MNRAS.480.3871C} used Gaussian process (GP) regression of principal components to construct a
multiwavelength surrogate calibrated to a fixed three-dimensional grid of  simulations \cite{2017Natur.551...80K}, describing flux $F_k$ from a single component of ejected material.  This study generated a ``two-component''
ejecta model by adding the fluxes of two independent calculations ($F=F_1+F_2$), ignoring any photon reprocessing effects.
More recently, Heinzel et al \cite{gwastro-mergers-em-CoughlinGPKilonova-2020}  applied this method to construct an anisotropic
surrogate depending on two components $M_{1},M_{2}$ and viewing angle, calibrating to their own anisotropic radiative transfer
calculations. They also included reprocessing effects,
showing that their previous simplified  approach which treats the radiation from each of the two components of the
outflow independently  introduces biases in inference for the components' parameters. 
These strong reprocessing or morphology-dependent effects are expected in kilonova light curves
 \cite{2020arXiv200400102K,2017ApJ...850L..37P, O_Connor_2020, O_Connor_2021}. %
Finally, a recent study by Breschi et al. \cite{2021arXiv210101201B} favored an anisotropic multicomponent model.

In this work, extending \cite{RisticThesis}, we apply an adaptive-learning technique to generate surrogate light
curves from  simulations of anisotropic kilonovae.  Starting with a subset of \nSimStart{} simulations reported in
\cite{kilonova-lanl-WollaegerNewGrid2020}, we use these adaptive learning methods to identify new
simulations to perform, refining our model with \nSimPlaced{}  simulations so far.  
We apply our surrogate light curves to reassess the parameters of GW170817.
We distribute the updated simulation archive, our current-best surrogate models, and our training algorithms at \texttt{https://github.com/markoris/surrogate\char`_kne}.

This paper is organized as follows.
In Section \ref{sec:Placement} we describe the kilonova simulation family we explore in this study and the active learning methods we
employ to target new simulations to perform.  We also briefly comment on our model's physical completeness.
In Section \ref{sec:Interpolation} we describe the specific procedures we employed to interpolate between our simulations
to construct surrogate light curves.
In Section \ref{sec:PE} we describe how we compare observations to our surrogate light curves to deduce the
(distribution of) best fitting two-component kilonova model parameters for a given event.  We specifically compare our
model to GW170817.
In Section \ref{sec:discussion} we describe how our surrogate models and active learning fit into the broader challenges
of interpreting kilonova observations.  
We conclude in Section \ref{sec:conclude}.

\section{Kilonova Simulation Placement}
\label{sec:Placement}

\subsection{Kilonova simulations}
\label{sec:kne_sims}

The kilonova simulations described in this work adopt a similar setup as and expand on the work of \cite{kilonova-lanl-WollaegerNewGrid2020}. 
The simulations discussed throughout were generated using the SuperNu \cite{2014ApJS..214...28W} time-dependent radiative transfer code, 
using tabulated binned opacities generated with the Los Alamos suite of atomic physics codes \cite{2015JPhB...48n4014F,2020MNRAS.493.4143F}.
We use results from the \textsc{WinNet} code \cite{2012ApJ...750L..22W} to determine radioactive heating and
composition effects.  We employ the thermalization model of  \cite{2016ApJ...829..110B}, but use a grey Monte Carlo
transport scheme for
gamma ray energy deposition \cite{2018MNRAS.478.3298W}.    

\begin{figure}
\includegraphics[width=\columnwidth]{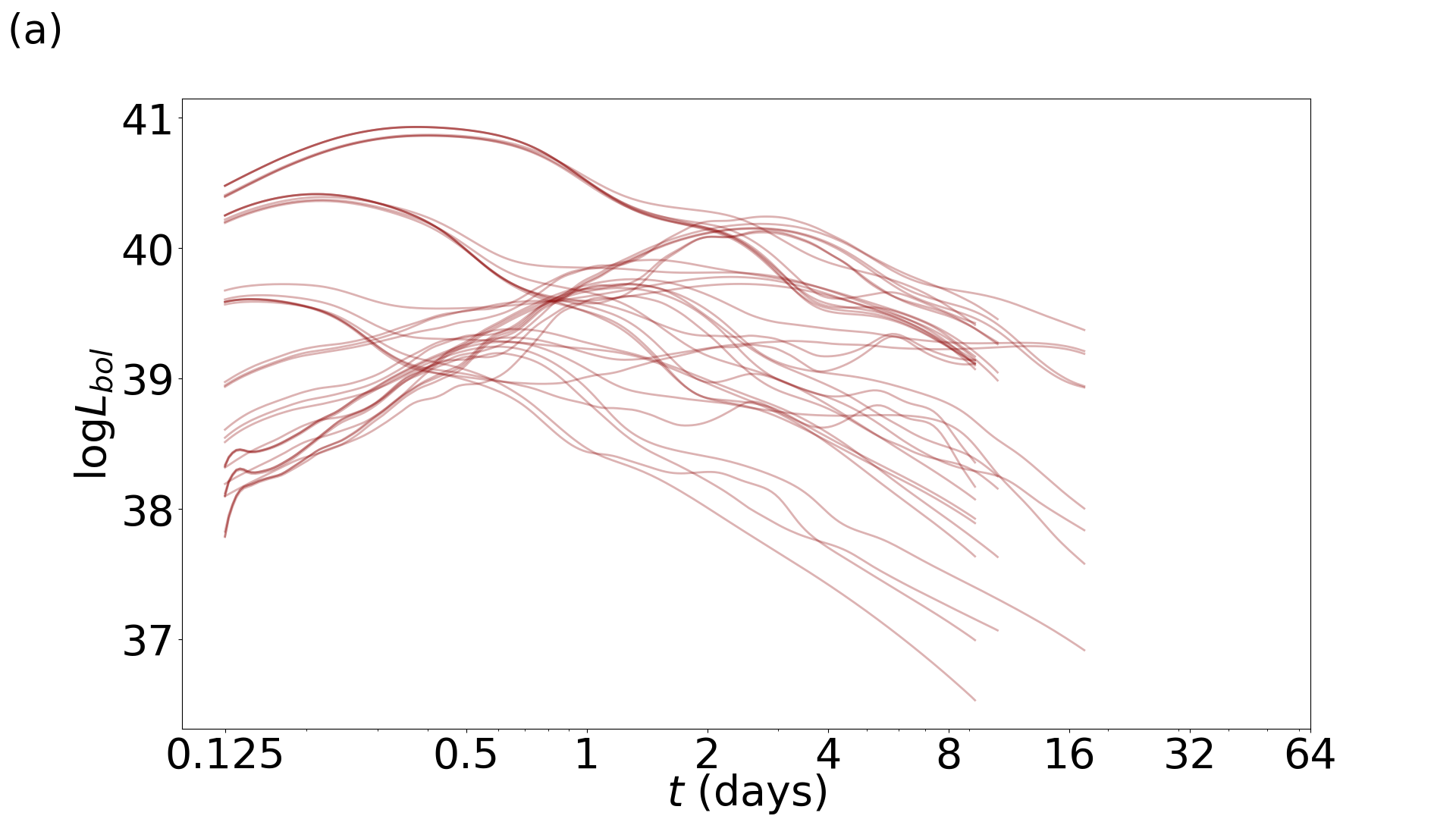}
\includegraphics[width=\columnwidth]{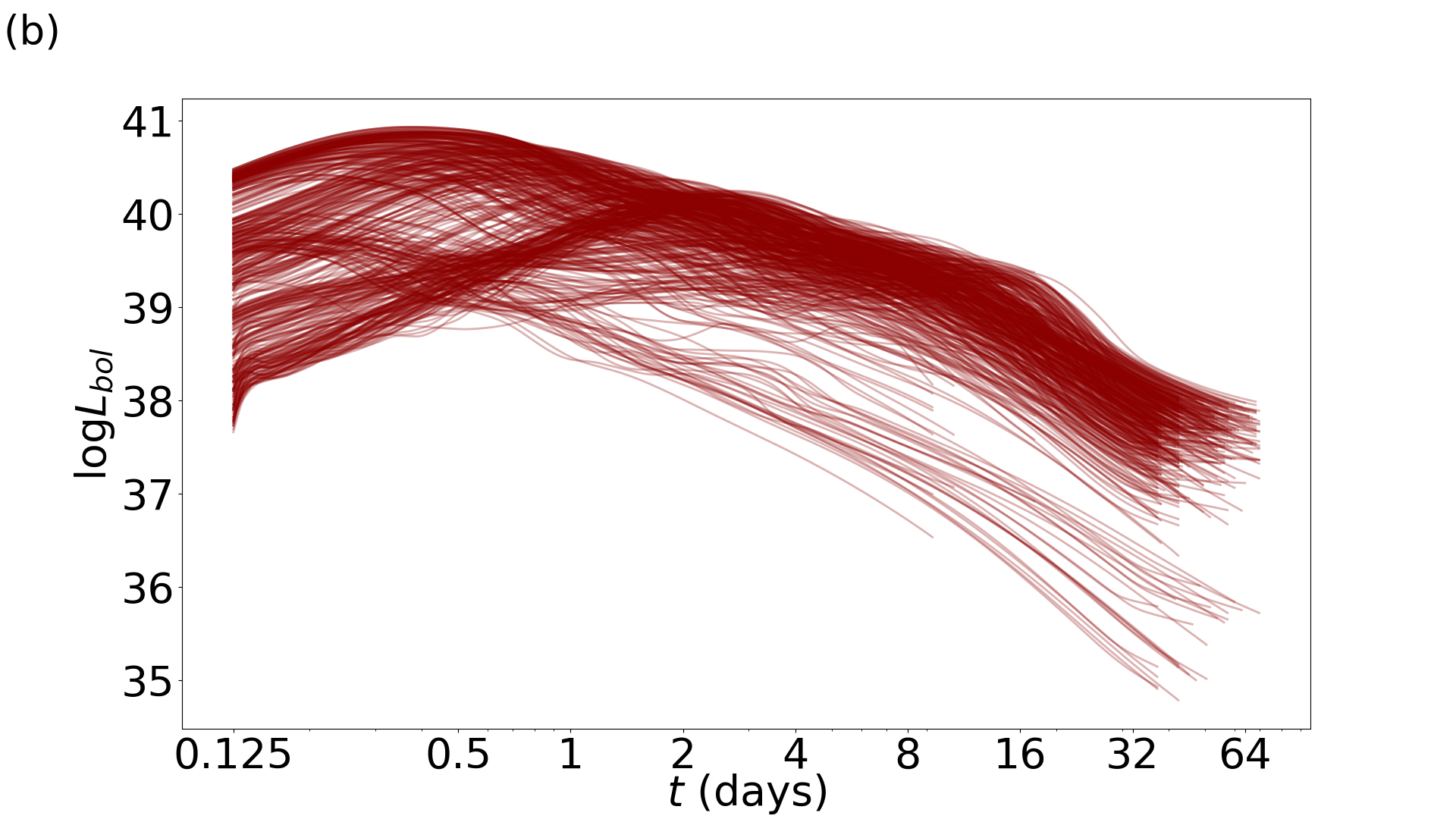}
\caption{\label{fig:sample_lc}\textbf{Bolometric luminosities of initial and adaptively placed simulations}: The top panel
shows the $\log_{10}$ bolometric luminosity in CGS units versus time in days for the simulations we initially used to train
our grid.  These simulations all extend out to roughly 8 days.   The bottom panel shows the bolometric light curves for our
adaptively placed simulations overlaid on top of the initial grid light curves.   Most of these simulations extend past 32 days.  Both panels exhibit
significant diversity in behavior and timescale.}
\end{figure}

The ejecta model is based on a symmetrically-shaped ideal fluid expanding in vacuum described by the
Euler equations of ideal hydrodynamics. The assumption of a radiation-dominated polytropic equation of state allows for
an analytic representation of the ejected mass $M$ and average velocity $\bar{v}$ as a function of
initial central density $\rho_0$, initial time $t_0$, and the velocity of the expansion front $v_{max}$
(Equations 11 and 12 in \cite{2018MNRAS.478.3298W}). When combined with Monte Carlo-based radiative transfer and a specified
elemental composition for the ejecta, the code produces time- and orientation-dependent spectra.  Convolving these spectra with
standard observational filters produces light curves such as the ones in Figures \ref{fig:sample_lc} and
\ref{fig:off_sample_interp}. %

Real neutron star mergers have (at least) two mechanisms for ejecting material, denoted as dynamical and wind ejecta \cite{2020ARNPS..7013120R}.
Due to the difference in formation mechanisms of dynamical and wind ejecta \cite{2019LRR....23....1M}, a multi-component
approach is necessary for accurate modeling. Each of the two types of ejecta, dynamical and wind, is modeled by a
separate component with a specified morphology, elemental composition, ejecta mass and ejecta
velocity. 
The components are modeled together as one radiative outflow \cite{2018MNRAS.478.3298W}.  The thermal decay energy is treated by mass-weighting
between the components where they overlap. %
The end product represents a time-dependent spectral energy distribution contained in 54 angular bins, equally spaced in
$\cos\theta$ from $1$ to $-1$. For the purposes of this study, the spectra are convolved with broadband filters to produce a series of
broadband light curves. Specifically, we use the LSST $grizy$ filters for optical and near-infrared bands, 2MASS $JHK$ filters
for longer wavelength near-infrared, and the mid-infrared $S$ filter for the Spitzer $4.5\;\mu$m band.
For each band and emission direction, we estimate the AB
magnitude for that filter, defined for a source at $10\unit{pc}$ in terms of the CGS energy flux $F_\nu$ per unit
frequency via
$
m_{X,AB} = -2.5 \log_{10}  \E{F_{\nu}} - 48.6
$.
All observations used in this work are provided or are translated into this  AB-magnitude system   \cite{1998A&A...333..231B,2007AJ....133..734B,2006MNRAS.367..454H}.
Because our simulations tend toward reflection-symmetric behavior across the $z=0$ plane, we only consider the independent information contained in the upper half ($z > 0$) of these angular bins. 
To reduce the acquisition cost of each simulation, we evolved each kilonova simulation in our initial grid out to $\tEndDays$
days.  To minimize data-handling and training cost, unless otherwise noted, we manipulate a subset of our simulation
output based on a log-uniform grid. For the initial simulations, this log-uniform grid consists of \nTimePoints{} time points ranging from $\tStartDays{}$ to $\tEndDays$ days.  
For the remaining simulations, this grid is extended in log-time to cover their available duration, up to
a maximum of 64 days.
Because of several systematics associated with modeling emission at early times (e.g., in the ionization states of the
medium and in the contribution from and interaction with any strong jet), we do not report on behavior prior to 3 hours
post-merger.
In this work, we use the orientation-averaged luminosity for simulation placement, but reconstruct the luminosity
continuously in angle and time.%

The original simulation hypercubes discussed in \cite{2018MNRAS.478.3298W,kilonova-lanl-WollaegerNewGrid2020} consider multiple wind ejecta morphologies and
compositions. To simplify the dimensionality of the problem, this work only considers simulations from the initial grid
with a peanut-shaped morphology \cite{2020arXiv200400102K}  and lower $Y_e = 0.27$ composition describing the wind ejecta. Table
\ref{tbl:grid_params} %
summarizes the parameters for the \nSimStart{} simulations in our four-dimensional  hypercube and highlights
variation in only ejected mass $M$ and average velocity $\bar{v}$ for each of the two components: the mass and velocity of the dynamical and wind
ejecta, denoted henceforth as $M_{d},v_d, M_{w},v_w$.  
Every simulation in our hypercube adopts the same morphologies for the dynamical and wind ejecta, respectively.
This initial simulation hypercube thus consists of only 2 of the 3 velocities and 3 of the 5 masses explored in the
companion study \cite{kilonova-lanl-WollaegerNewGrid2020}.
\begin{table}[h!]
\begin{center}
\begin{tabular}{|c@{\hskip 2mm}c@{\hskip2mm}c@{\hskip 5mm}c@{\hskip5mm}c|} 
\hline
Ejecta & Morphology & $Y_e$ & $M\textsubscript{ej}$ & $\bar{v}$ \\ [0.5ex] 
 &  &  & $M_\odot$ & $c$ \\ [0.5ex] 
\hline\hline
Dynamical & Torus & $0.04$ & \begin{tabular}{@{}c@{}} 0.001, 0.01, 0.1  \end{tabular} & \begin{tabular}{@{}c@{}} 0.05,  0.3  \end{tabular} \\\hline
Wind & Peanut & $0.27$ & \begin{tabular}{@{}c@{}} 0.001, 0.01,  0.1  \end{tabular} & \begin{tabular}{@{}c@{}} 0.05,  0.3  \end{tabular} \\ %
\hline
\end{tabular}
\end{center}
\caption{\textbf{Kilonova simulation parameters}: Within the framework of models explored in
  \cite{kilonova-lanl-WollaegerNewGrid2020}, parameters of the initial kilonova simulations used to initialize our
  adaptive learning process in this work.  All simulations used in this work adopt a two-component model where the
  morphology and composition of  each component is fixed. }
\label{tbl:grid_params}
\end{table}
As expected and  discussed elsewhere \cite{kilonova-lanl-WollaegerNewGrid2020}, these simulations exhibit significant viewing-angle
dependence on the relative speed of the components.  %
The obscuration of the wind by the dynamical ejecta becomes less significant closer to the symmetry axis and the 
peanut morphology itself also produces orientation dependence.
The two-component model shows ``blanketing'' of slow
blue components by fast red components \cite{2015MNRAS.450.1777K}.
Also expected and observed are qualitative trends versus the component masses and velocities: more wind ejecta mass
increases the $g$-band luminosity along the symmetry axis.

\subsection*{Illustrating systematics of kilonova simulations}
Before extensively discussing our ability to reproduce this specific family of simulations, we first comment on their
systematic limitations.  Our simulation archive explores only a limited range of initial conditions for the ejecta, with specific assumptions
about the composition, morphology, and velocity profiles; with specific assumptions about nucleosynthetic heating; and
with specific assumptions about (the absence of) additional power and components, such as a jet or a central source to
provide additional power or light \cite{2021MNRAS.500.1772N,2021MNRAS.502..865K, Piro_2018, Ai_2018}.
Several previous studies have indicated that these and other aspects of kilonova simulations can noticably impact the
outcome
\cite{2018MNRAS.478.3298W,2020ApJ...899...24E,2019ApJ...880...22W,2020arXiv201214711K,2021ApJ...906...94Z,2021ApJ...910..116K,2017ApJ...850L..37P}.  Where possible, we very briefly comment on how current and previous SuperNu
simulations' results change when making similar changes in assumptions.    

\begin{figure}
\includegraphics[width=\columnwidth]{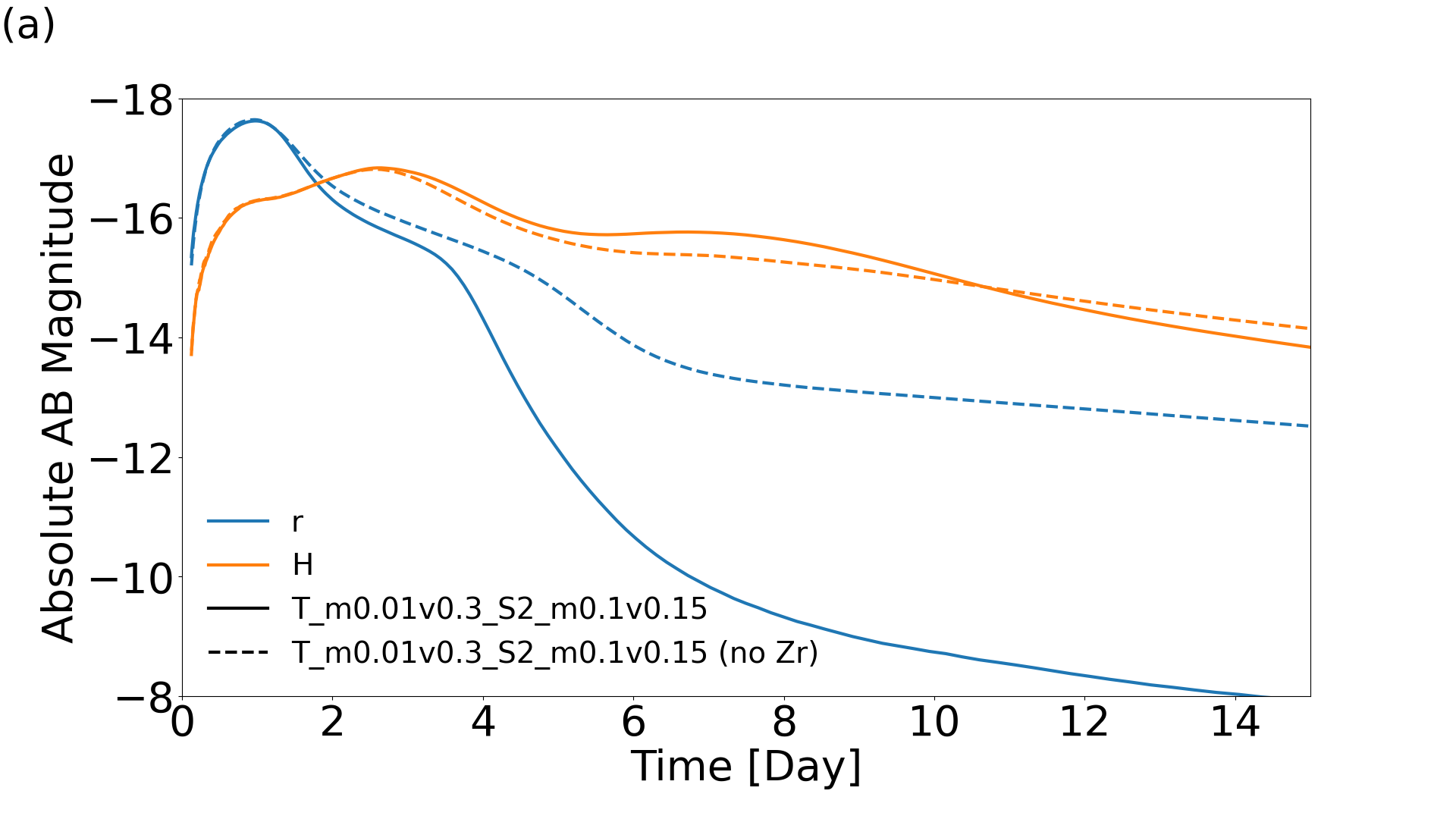}
\includegraphics[width=\columnwidth]{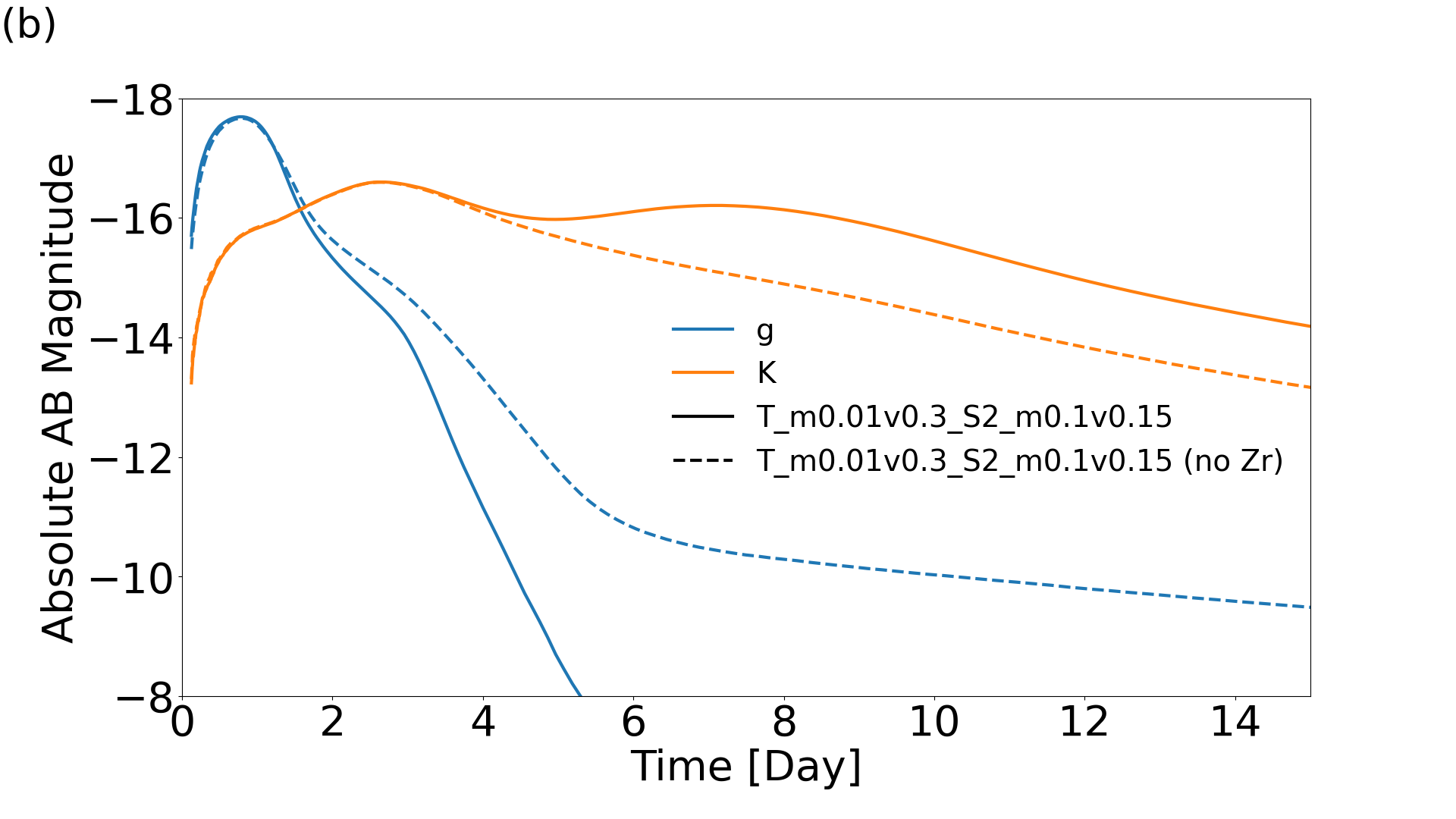}
\caption{\label{fig:Zr}\textbf{Impact of removing Zirconium}: Solid and dashed lines show simulations with otherwise identical
  assumptions about composition, morphology, and velocity structure, differing only by the presence (solid) and
  elimination (dashed) of Zr. The selected simulation parameters, $M_{d}=0.01 M_{\odot}$, $v_d=0.3 c$, $M_{w}=0.01 M_{\odot}$,
and $v_w=0.15 c$, are our closest-matching representation of the simulation parameters considered
during the Zr-omitting study in \cite{2020arXiv201214711K}.
}
\end{figure}

Prior work with SuperNu has explored the impact of composition \cite{2020ApJ...899...24E}.  
However, recently, Kawaguchi et al 2020 \cite{2020arXiv201214711K} (henceforth K20) demonstrated that Zr makes a substantial contribution to the final light
curve. Figure  \ref{fig:Zr} shows how our simulations depend on a similar change in composition, noting substantial
change in the late-time optical light curves when we remove Zr.

As demonstrated by many previous studies using SuperNu, the morphology and velocity structure also has a notable impact on the post-day light curve behavior
\cite{2018MNRAS.478.3298W,2021ApJ...910..116K,2017ApJ...850L..37P}.   
Several other groups have demonstrated similar strong morphology and orientation dependence in their work \cite{2020ApJ...897..150D,2020arXiv201214711K,2021arXiv210101201B,gwastro-mergers-em-CoughlinGPKilonova-2020}. 
For example,   in their Figure 8, K20 demonstrate how the light curve changes when a specific polar component of the
ejecta is removed.
Uncertain nuclear physics inputs also propagate into notable uncertainties about the expected light curve; see, e.g., \cite{2021ApJ...906...94Z,2020arXiv201011182B}.
Even for the same morphology and amount of ejecta, nuclear physics uncertainties can modify the effective heating rate,
particularly for material with low $Y_e$ which has the greatest prospect for producing r-process elements.

Given limited exploration of possible kilonova initial conditions and physics, we can only at present quantify the
uncertainties of the type listed above.  In future work, we will employ our parameterized models   to assess the impact of
these uncertainties on inferences about kilonova parameters.   Future work could require kilonova models which include
EOS parameters to enable  joint inference which also
simultaneously constrains the equation of state.

As discussed elsewhere \cite{kilonova-lanl-WollaegerNewGrid2020}, at late times some light curves show a modest deficit of blue light ($g$-band) 
relative to observations of GW170817 (unless the dynamical ejecta mass is large).  Notably, our $g$-band light curves fall off significantly more rapidly after
their peak in all viewing directions and for most parameters considered here. 
Previous work with other morphologies also recovers similar falloff in these bands,
see e.g. \cite{tanvir17}, though additional components could conceivably contribute. 
 Similar $g$-band behavior has  been seen in other
detailed kilonova simulations; see, e.g, Figure 12 in \cite{2020ApJ...889..171K}.
As noted above, this behavior depends on the assumed composition, notably Zr.  

\subsection{Interpolation Methodology}
\label{sec:interp_method}

In this work, we principally interpolate using Gaussian process (GP) regression.  In GP regression, given
training data pairs $(x_a,y_a)$, the estimated function $\hat{y}(x)$ and its variance $s(x)^2$  are approximated by
\begin{subequations}
\label{eq:gp}
\begin{align}
\hat{y}(x) &= \sum_{a,a'}k(x,x_a) K^{-1}_{aa'}y_{a'} \\
s(x)^2 &= k(x,x) - k(x,x_a)K^{-1}_{aa'} k(x_{a'},x)
\end{align}
\end{subequations}
where the matrix $K_{aa'} = k(x_a,x_{a'})$ and where the function $k(x,x')$ is called the kernel of the Gaussian
process.  In this work, unless otherwise noted, we used a squared-exponential kernel and a white noise (diagonal) kernel
\begin{eqnarray}
k(x,x') =  \sigma_o^2 e^{-(x-x')Q(x-x')/2} + \sigma_n^2 \delta_{x,x'}
\end{eqnarray}
where $Q$ is a diagonal matrix of possible length scales and $\sigma_0,\sigma_n$ are hyperparameters that characterize
the amount of noise allowed in the problem. 
The other interpolation method considered in this work was random forest (RF) regression \cite{breiman2001}. Unlike the GP, the 
RF output had no error quantification and was used primarily as a consistency check on the Gaussian process
prediction. 
Unless otherwise noted, we performed all GP and RF regression with \textsc{scikit-learn} \cite{scikit-learn}.

Because of the substantial dynamic range of our many outputs, we interpolate the $\log_{10}$ luminosity (for
placement) or AB magnitudes (for all other results).
Unless otherwise noted, we quantify the performance of our interpolation with the RMS difference between our prediction
and the true value
\begin{equation}
\ell^2 = \frac{1}{n} \sum_{j=1}^{n} (y_j - \log_{10}(L_\text{bol})_j)^2,
    \label{eq:simple_loss}
\end{equation}
[This expression overweights the importance of large errors when the source is not detectable at late times; see Appendix
\ref{ap:validate_pe}].

We employ GP interpolation in two standard use cases.   In the first case, used for our exported production results, we interpolate the AB magnitude $
m_\alpha(t_*|\Lambda)$ at some fixed reference time $t_*$ and band $\alpha$ versus our four simulation  hyperparameters
 (and, in the end, also across the extrinsic parameters of angle and wavelength) contained in $\Lambda$.  In this case, the prediction $y(x_a)$ has a single scalar value at each point; the
$x_a$ refer to model hyperparameters; and the interpolation provides us with a scalar function of four or more
variables.  GP regression [Eq. (\ref{eq:gp})] provides an error estimate for $m_\alpha$ at this specific time
$t_*$, which in general will depend on time.
In the second case, used for simulation placement, we interpolate the log bolometric luminosity
\emph{light curve} $\log_{10} L_{bol}(t|\Lambda)$ versus \emph{all time}.  
[In terms of each simulation's spectrum, the bolometric luminosity is
$
L_{bol} = 4 \pi R^2 \int_0^{\infty} F_{\nu}d\nu
$ 
where R = 10 pc.]
In this case, the prediction $\vec{y}(x_a)$ is vector-valued at each point; the $x_a$ refer to model
hyperparameters; and the interpolation provides us with a vector-valued function of four or more variables.
For simplicity and given our use case, we reduce our error estimate to a single overall value for the entire light curve, reflecting the overall
uncertainty in $\vec{y}(x_a)$.

\subsection{Active Learning Scheme}
\label{sec:active_learning}

Gaussian processes have long been used for active learning because they provide an error estimate: follow-up simulations
can be targeted in regions with the largest expected error (and thus improvement)  \cite{book-Murphy-MachineLearning}.
We follow this approach in our active learning scheme; see \cite{ZacksSolomon1970,krause07nonmyopic,Cohn1996,Gal2017,MacKay92bayesianmethods,Srinivas10,Mockus78,Wu16} for a broader discussion of active
learning methods and their tradeoffs. 
To reduce the data volume needed for targeting followup simulations, we used vector-valued interpolation as described
above, applied to \emph{orientation-averaged outputs} of our simulations.  This
approach has the substantial advantage of  providing a single error estimate per light curve (both in training and off-sample), which we can immediately
use as an objective function in a minimization algorithm.

We pursued an active learning simulation placement approach in order to maximally explore the parameter space and reduce
the amount of redundant information obtained from each new simulation. The subset of \nSimStart{}  light curves discussed in Section
\ref{sec:kne_sims} was used as the initial training set.
Thousands of parameter combinations were subsequently drawn from uniform distributions with maxima and minima matching
those of the varied parameters in Table \ref{tbl:grid_params}. Each of these parameter combinations was evaluated by an
interpolator to produce an initial light-curve prediction as well as an error on the entire light-curve output. The prediction with
the largest error across all the tested parameter combinations was selected as the next placed simulation.

\begin{figure}
\includegraphics[width=\columnwidth]{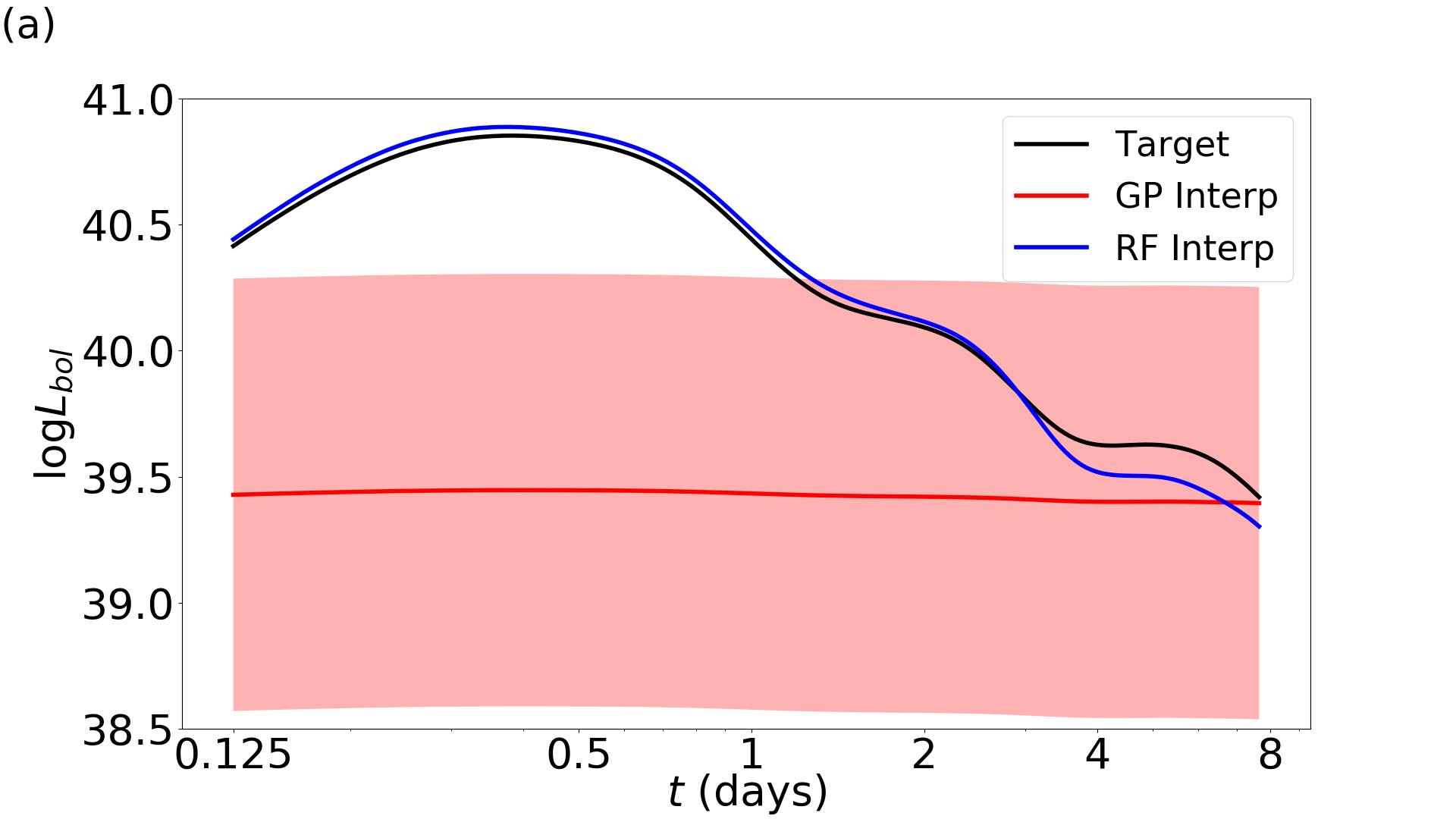}
\includegraphics[width=\columnwidth]{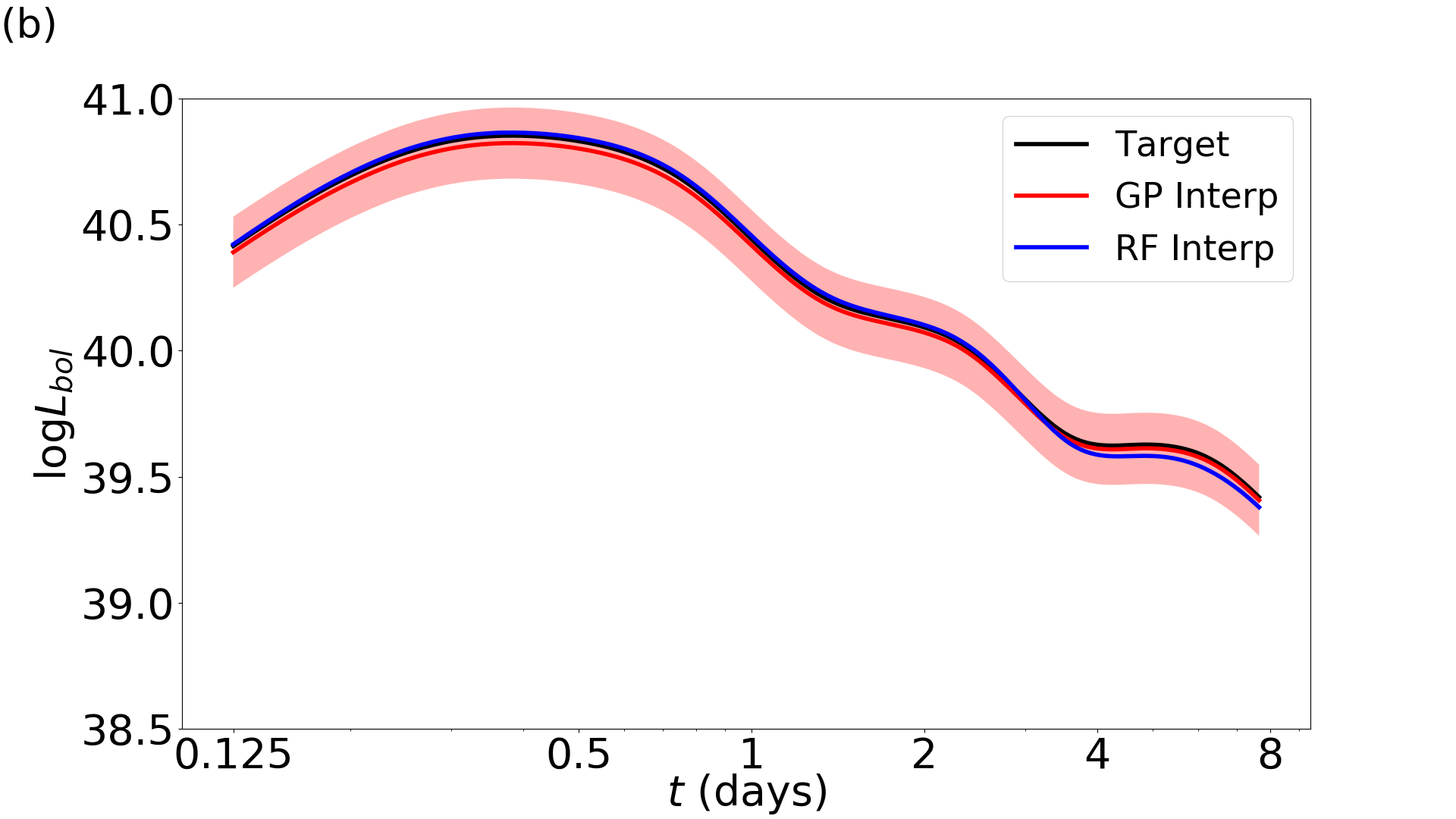}
\caption{\label{fig:GP_before_after} \textbf{Impact of adaptive placement on interpolation:}
 Example  of interpolation output at a point with large predicted fitting error, both before and after placing the
 simulation.  In both panels, the solid black curve shows the true simulated bolometric light curve versus time.
The red band shows the GP-predicted one-sigma error bar about the expected value.
 \emph{Top panel}:  Predictions from our RF and GP interpolations versus time.  The large error and low practical
 utility of the GP fit is apparent.  \emph{Bottom panel}: After including this simulation in the training set, the revised RF and
 GP predictions much more closely conform with this specific simulation as expected.
}
\end{figure}

\subsection{Prediction Improvement and Interpolation Results}

We verified  our active learning strategy for simulation  placement by   randomly sampling  combinations of
parameters and creating two light curve predictions based on those parameters. The first prediction was trained solely on our
initial grid of simulations from Section \ref{sec:kne_sims}, while the second prediction was trained on the same initial
grid, but with an added simulation output characterized by the aforementioned random combination of parameters. Figure
\ref{fig:GP_before_after} shows these before- and after-inclusion predictions which show that, as expected, the GP
interpolation capability is improved. 
This pair of figures anecdotally illustrates the degree to which new training data improves our surrogate light curve models.

With over 400 placed simulations since the start of the active learning process, the training library is built up
enough to allow for physically meaningful interpolation of off-sample events.   The performance of our adaptive learning
is best illustrated with our production-quality interpolation scheme, illustrated in Figure \ref{fig:off_sample_interp}
and described in the next section.

Despite producing many follow-up simulations, we achieve success with a
very  sparse coverage of our parameter space.  To illustrate the sparsity of our parameter space coverage, and how slowly
our added simulations increase coverage, we evaluated the median ``inter-simulation'' distance,
using a simple Euclidean ($L^2$) norm over $\log_{10} L_{bol}(t_k)$ for several reference times $t_k$. 
As expected given the high apparent dimension of our output, this median distance changes very
slowly with  $n$, owing to the large effective dimension of the output light curves.   The median distance is also
 larger than the residual error in our fit, as reported below.  The success of our interpolation relies not on an
overwhelmingly large training sample, but on the smoothness and predictability of our physics-based light curves.

\section{Light curve interpolation}
\label{sec:Interpolation}

\subsection{Stitched fixed-time interpolation}
\begin{figure}
\includegraphics[width=\columnwidth]{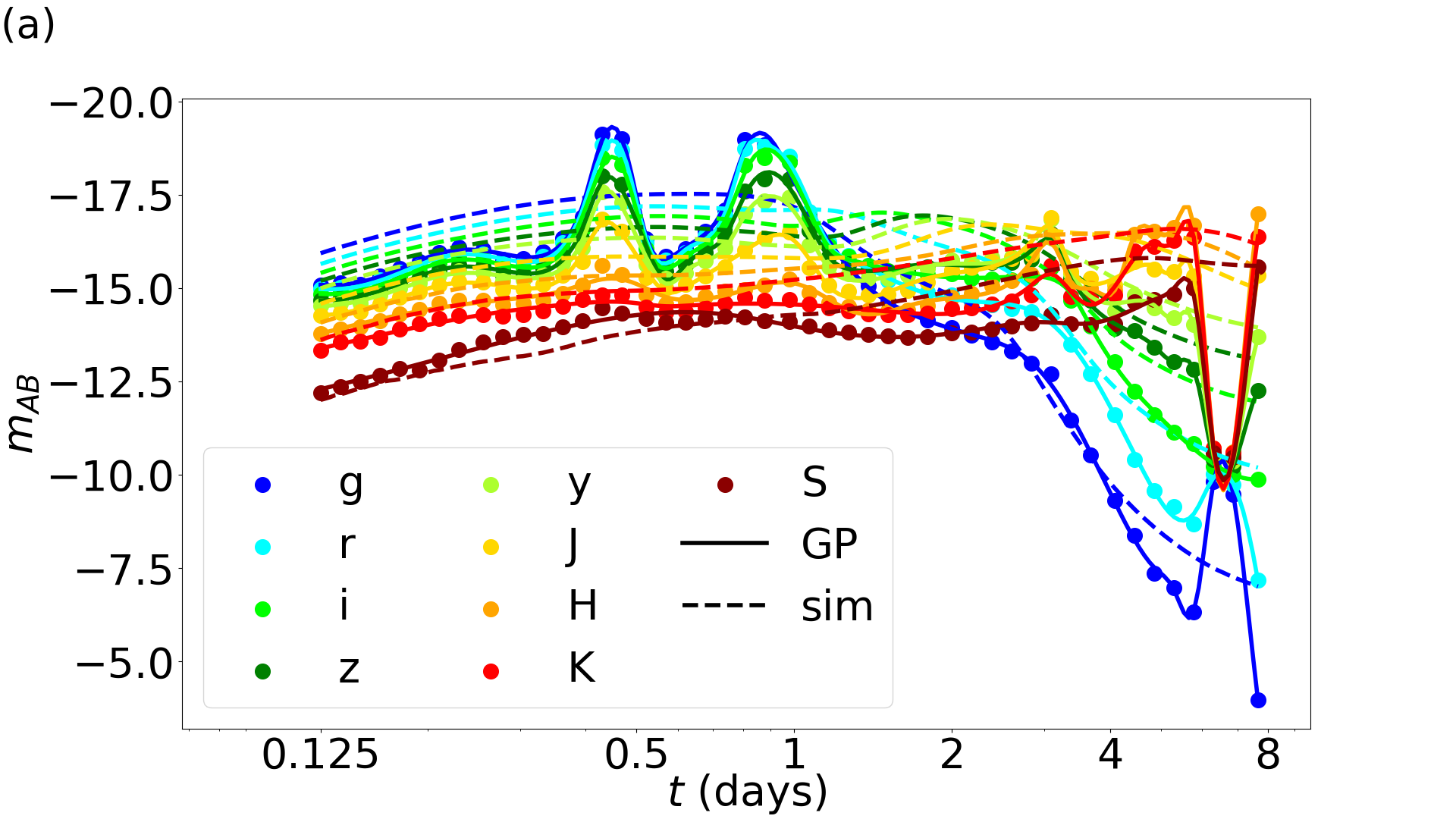}
\includegraphics[width=\columnwidth]{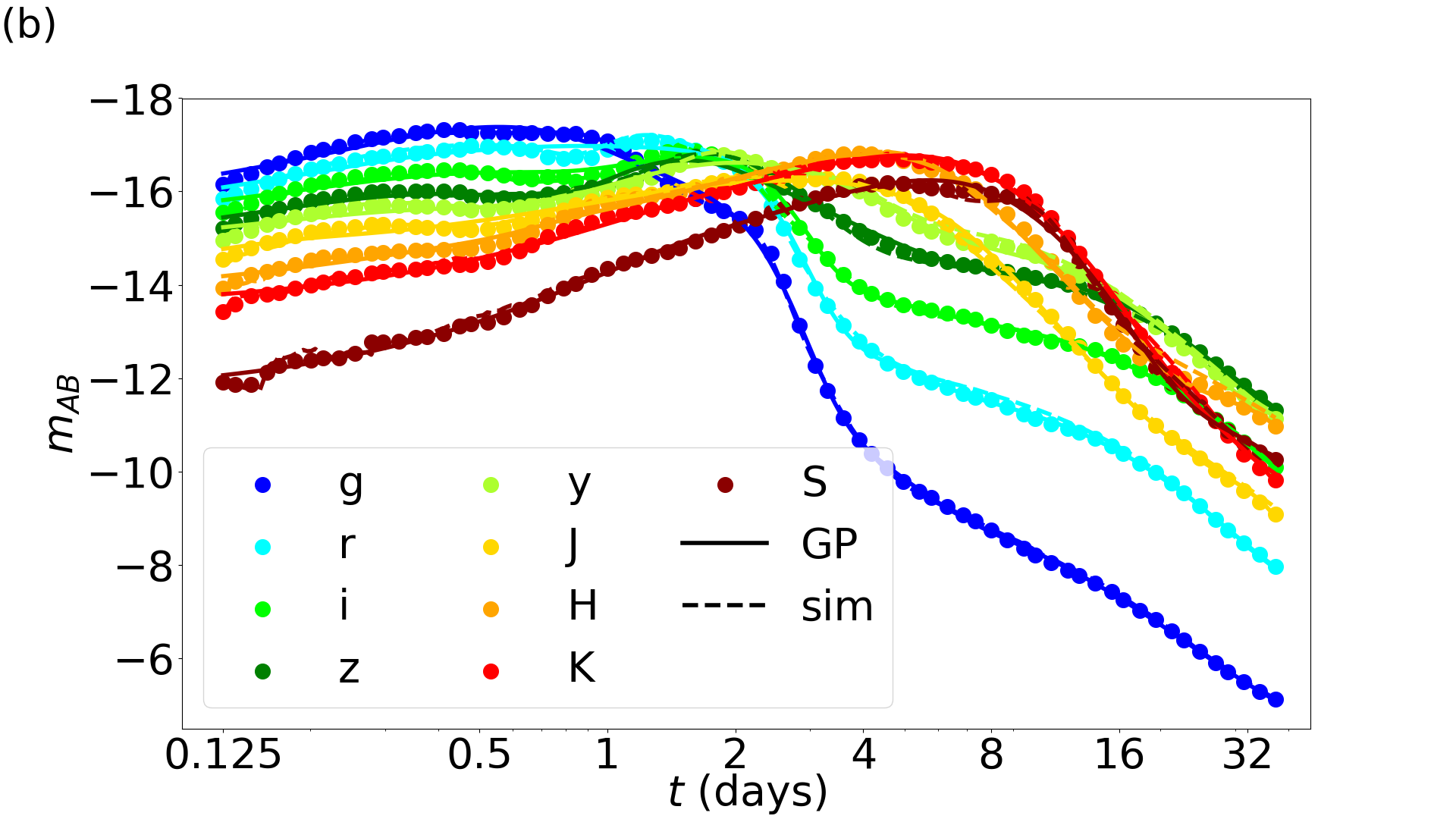}
\caption{\textbf{Off-sample interpolation with original and refined grid: } Example of an interpolated stitched fixed-time prediction compared to a simulation
output created from the same corresponding input parameters.  The top panel shows our estimate based on the initial
$\nSimStart$ simulations; the bottom panel shows the result after adaptive learning.   Different colors denote different filter bands, described
in the legend.  The dashed lines show full simulation output for each band.   The colored points show our interpolated
bolometric magnitude predictions at the $\nTimePoints$ evaluation times.  The solid lines show our final
interpolated light curves, interpolating between the points shown.   The largest error in this example occurs for the
$g$-band at late times.  The simulated parameters and viewing angle for this configuration are $M_{d}=0.097050 M_{\odot}$, $M_{w}=0.083748 M_{\odot}$,
$v_d=0.197642 c$ and $v_w=0.297978 c$, viewed on axis ($\theta=0$). The exaggerated modulations in the top panel's solid
lines and dotted curves illustrate interpolation failures, arising from adopting an initially insufficient training set.}
\label{fig:off_sample_interp}
\end{figure}

To efficiently interpolate across the whole model space, we follow a strategy illustrated in Figure 1 of
\cite{2014PhRvX...4c1006F}: we pick several fiducial reference times $t_{q}$ (and angles);  use GP interpolation to produce an
estimate $m_\alpha(t_q|\Lambda)$ versus $\Lambda$;  interpolate in time to construct a continuous
light curve at the model hyperparameters $\Lambda$ at each reference angle; and then interpolate in angle to construct a
light curve for an arbitrary orientation.    For an error estimate, we stitch together the error estimates in
each band to produce a continuous function of time.  
Figure  \ref{fig:off_sample_interp} shows the output of our interpolation (smooth lines), compared to a validation
simulation at the same parameters (dashed lines).   Our predictions generally agree, though less so for the shortest wavelengths at the
latest times.   
Subsequent figures also illustrate the typical GP error estimate, which is usually $O(0.1)$ in $\log_{10} L$ for most bands
and times considered. 

\subsection{Trends identified with interpolated light curves}
\begin{figure}
\includegraphics[width=0.925\columnwidth]{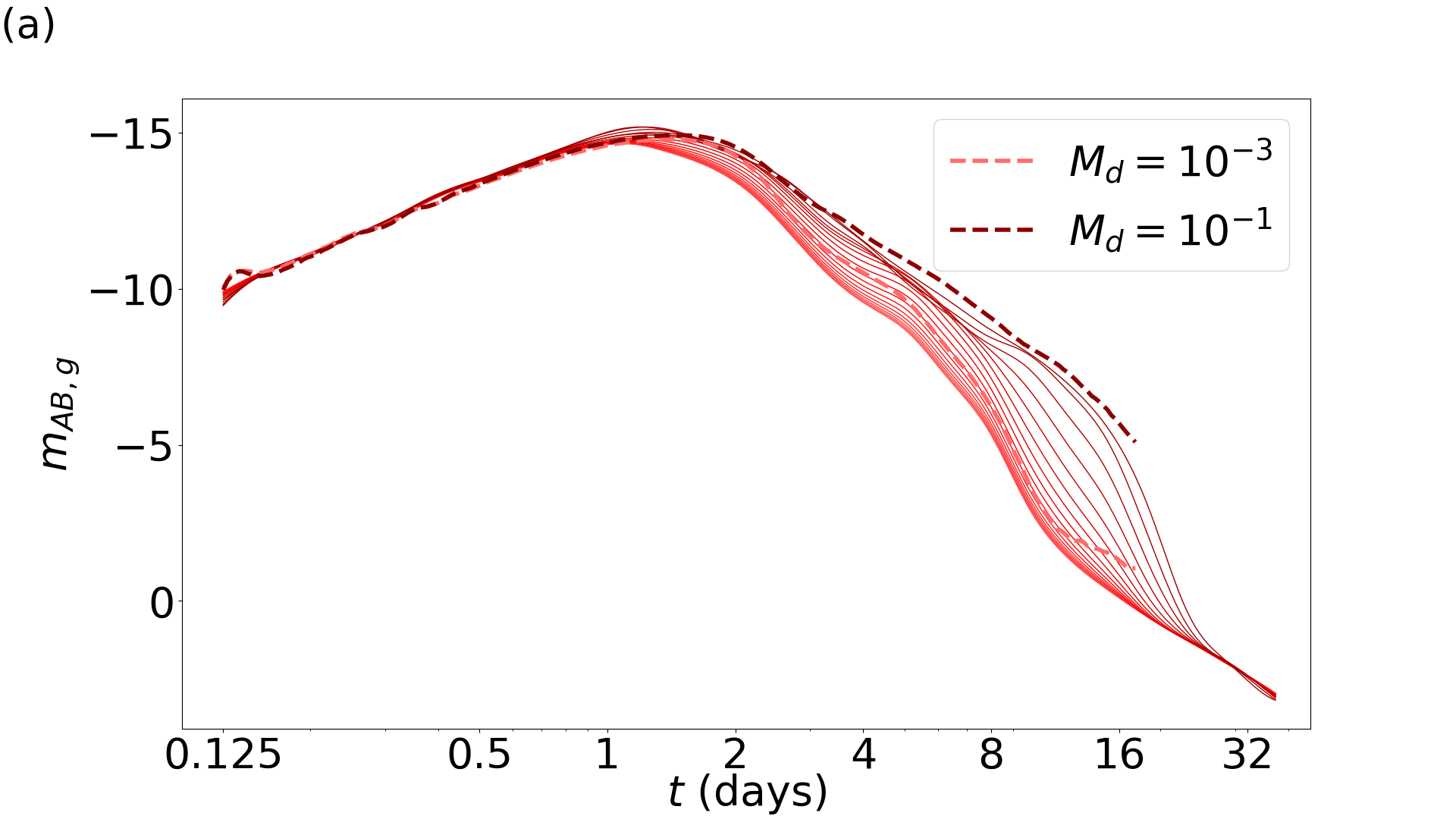}
\includegraphics[width=0.925\columnwidth]{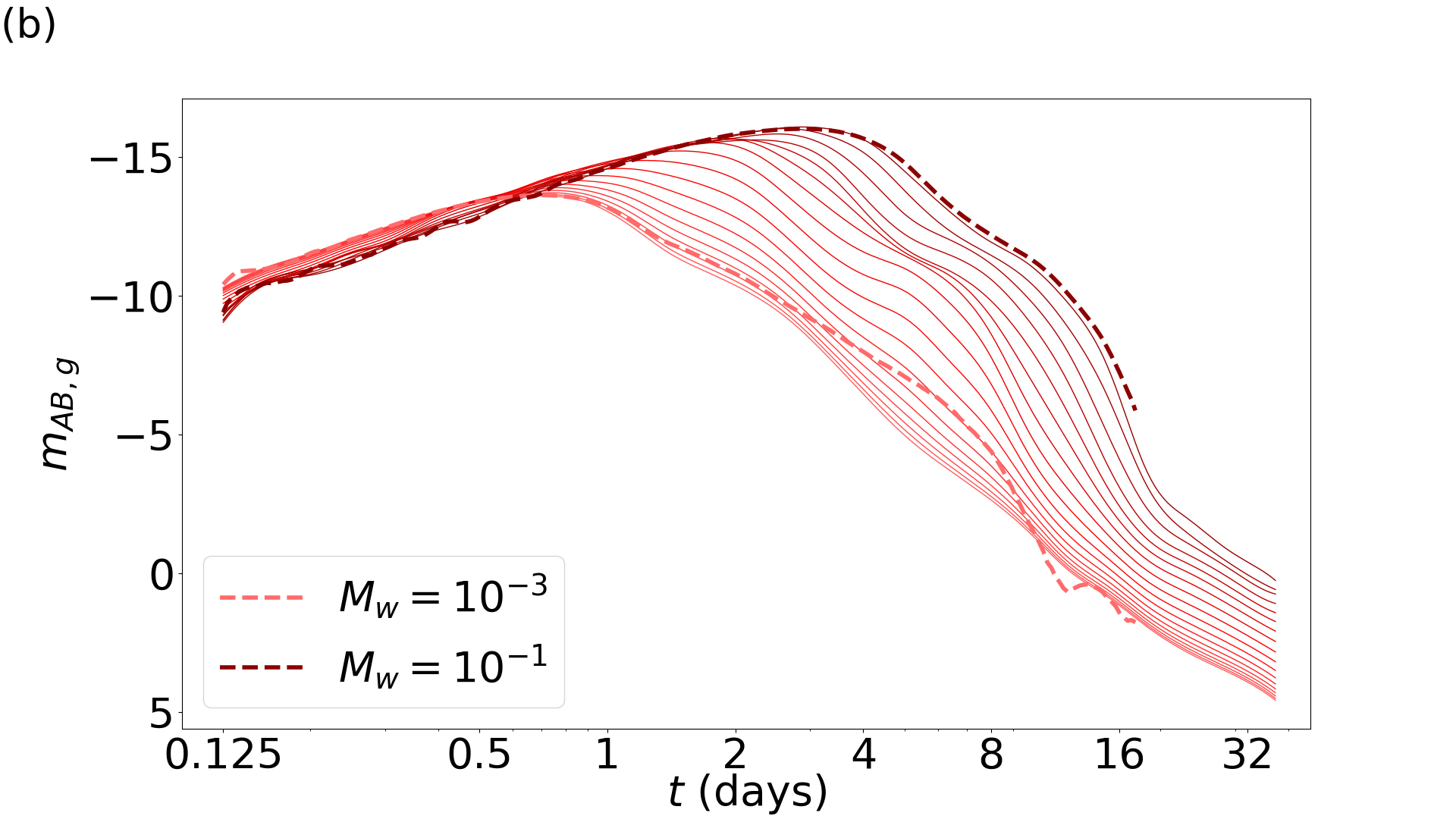}
\includegraphics[width=0.925\columnwidth]{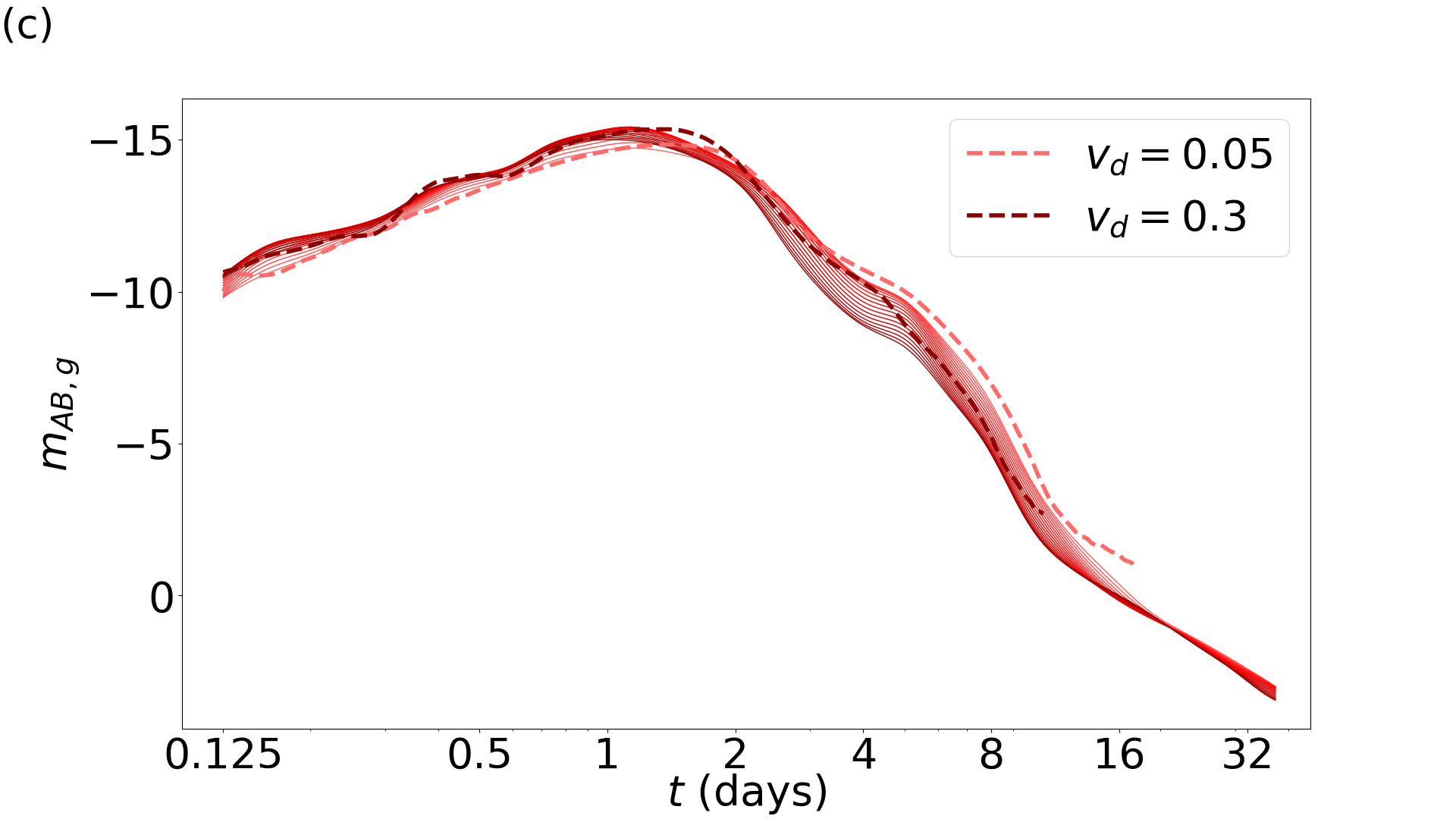}
\includegraphics[width=0.925\columnwidth]{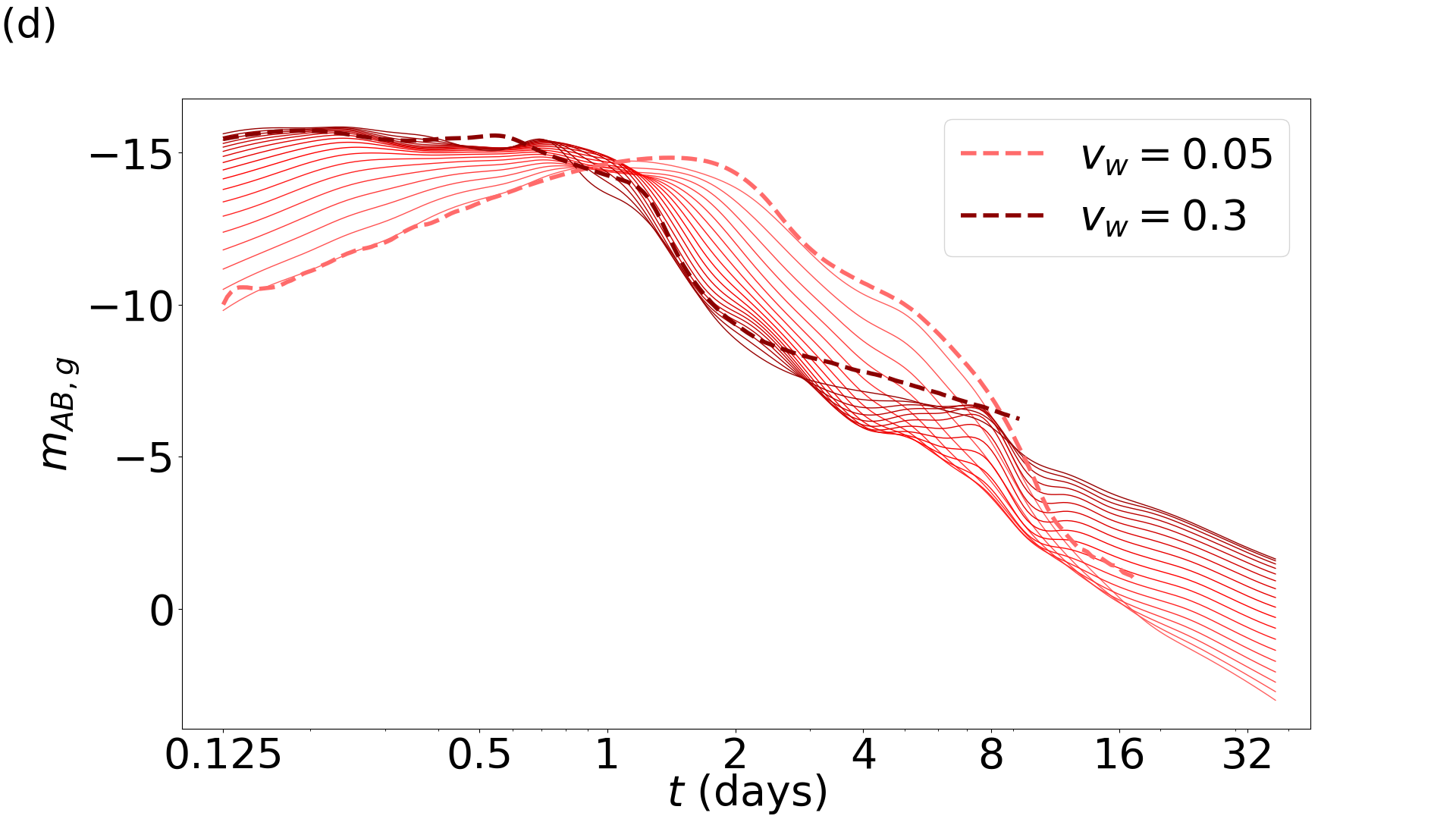}
\caption{\label{fig:CharacterizeTrends:OneParameter}\textbf{Interpolated and simulated $g$-band light curves}:
 In this figure, we generate $\log L_g(t|\Lambda)$ for a
  one-parameter family of simulations $\Lambda$ where either one of the $M$ parameters vary from $0.001 M_\odot$ to $0.1
  M_\odot$ or one of the $v$ parameters vary from $0.05c$ to $0.3c$, and the viewing angle is $\theta=0$. 
  The remaining model parameters are fixed to $(M/M_\odot,v/$c$) =  (0.01,0.05)$.  Contours in $M$
  are uniform in $\log M$, while those for $v$ are linearly uniform.  For
  comparison, the heavy dashed lines show the initial training simulation results for the two parameter endpoints.
The $g$-band light curve has the largest dynamic range and is the most sensitive to interpolation errors; notably, the
interpolation does not always conform tightly to the underlying simulation data at late times.
}
\end{figure}

In Figure \ref{fig:CharacterizeTrends:OneParameter} we show the results of our fit evaluated at a fixed viewing angle ($\theta=0$), varying one parameter at a time
continuously, relative to a fiducial configuration with $M_{d}=M_{w}=0.01 M_\odot$, $v_w/c=v_d/c=0.05$.
The fixed value for the ejected mass of $M=0.01 M_\odot$ was chosen as the middle ground of the initial grid's sampled mass space, which
does not introduce any biases toward lighter or heavier masses. Since no similar central value was initially available for the velocity
parameters, the lower value was selected in the case of both components. The slower velocity resulted in the ejecta not dissipating
as quickly and allowed for more variation in the light curves as the non-static parameter was varied. 
For this viewing angle, changes in the amount and velocity of the dynamical ejecta have relatively modest effect,
in large part because that ejecta is concentrated in the equatorial plane.  By contrast, changes in the mostly polar
wind ejecta has a much more substantial impact on the polar light curve ($\theta=0$). 
Specifically, increasing the amount of wind ejecta brightens and broadens the light curve, as expected from classic
analytic arguments pertaining to how much material the light must diffuse through \cite{1980ApJ...237..541A,1982ApJ...253..785A,Chatzopoulos_2012,2019LRR....23....1M}.
Similarly, increasing the velocity of wind ejecta causes the peak to occur at earlier times (diffusion is easier) and be
brighter.

\subsection{Interpolation in viewing angle}

\begin{figure}[ht!]
\includegraphics[width=0.925\columnwidth]{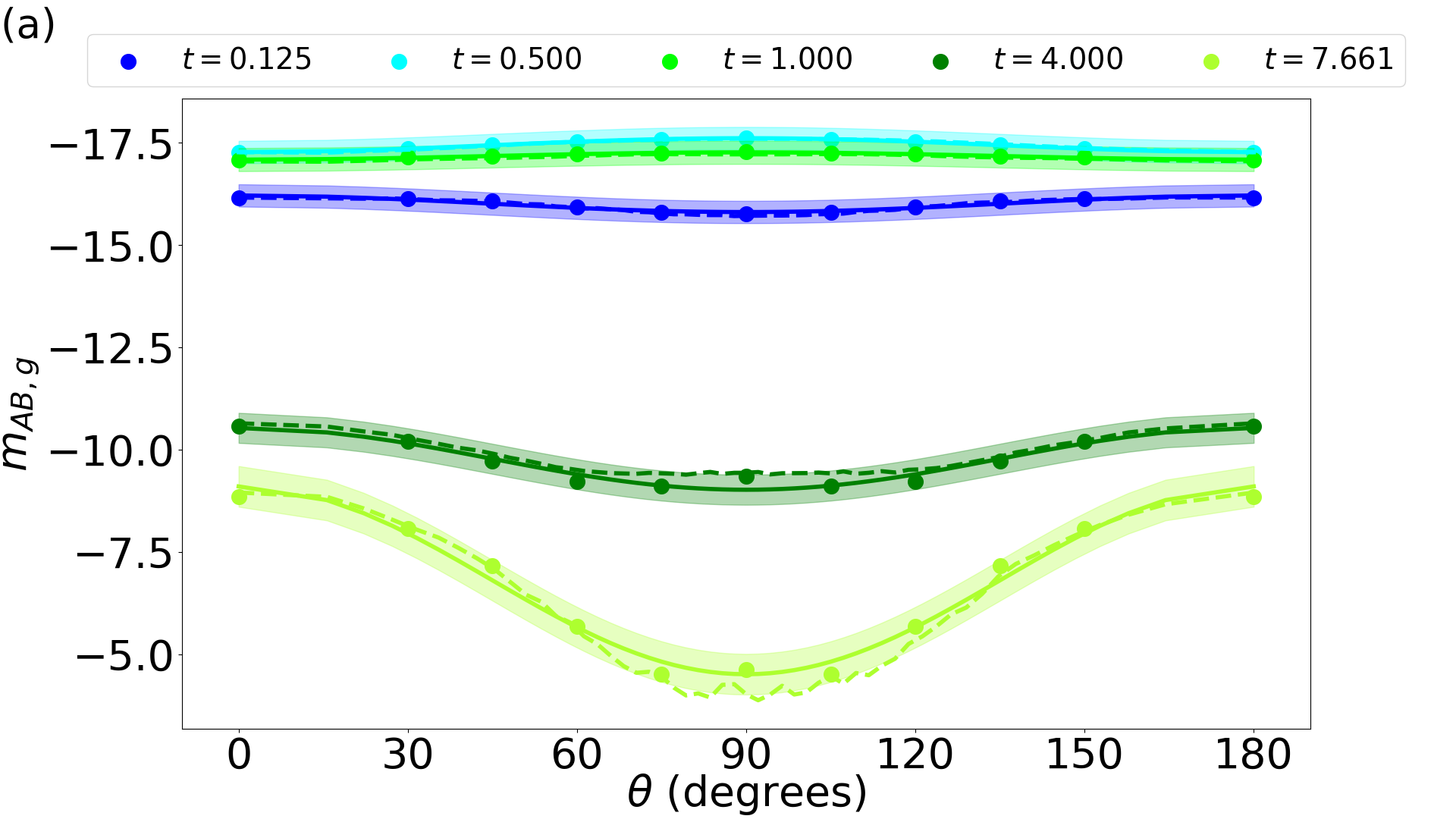}
\includegraphics[width=0.925\columnwidth]{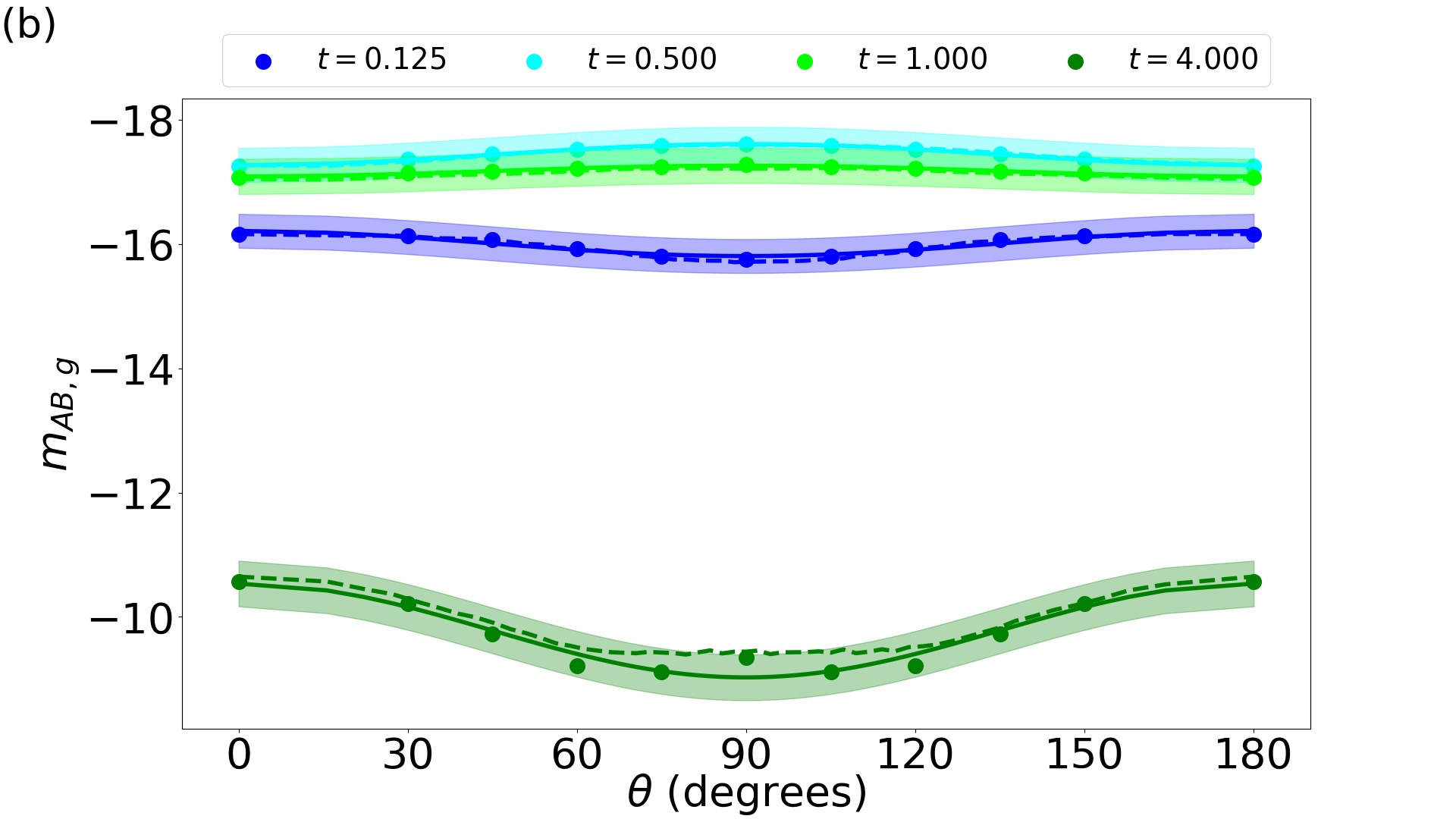}
\includegraphics[width=0.925\columnwidth]{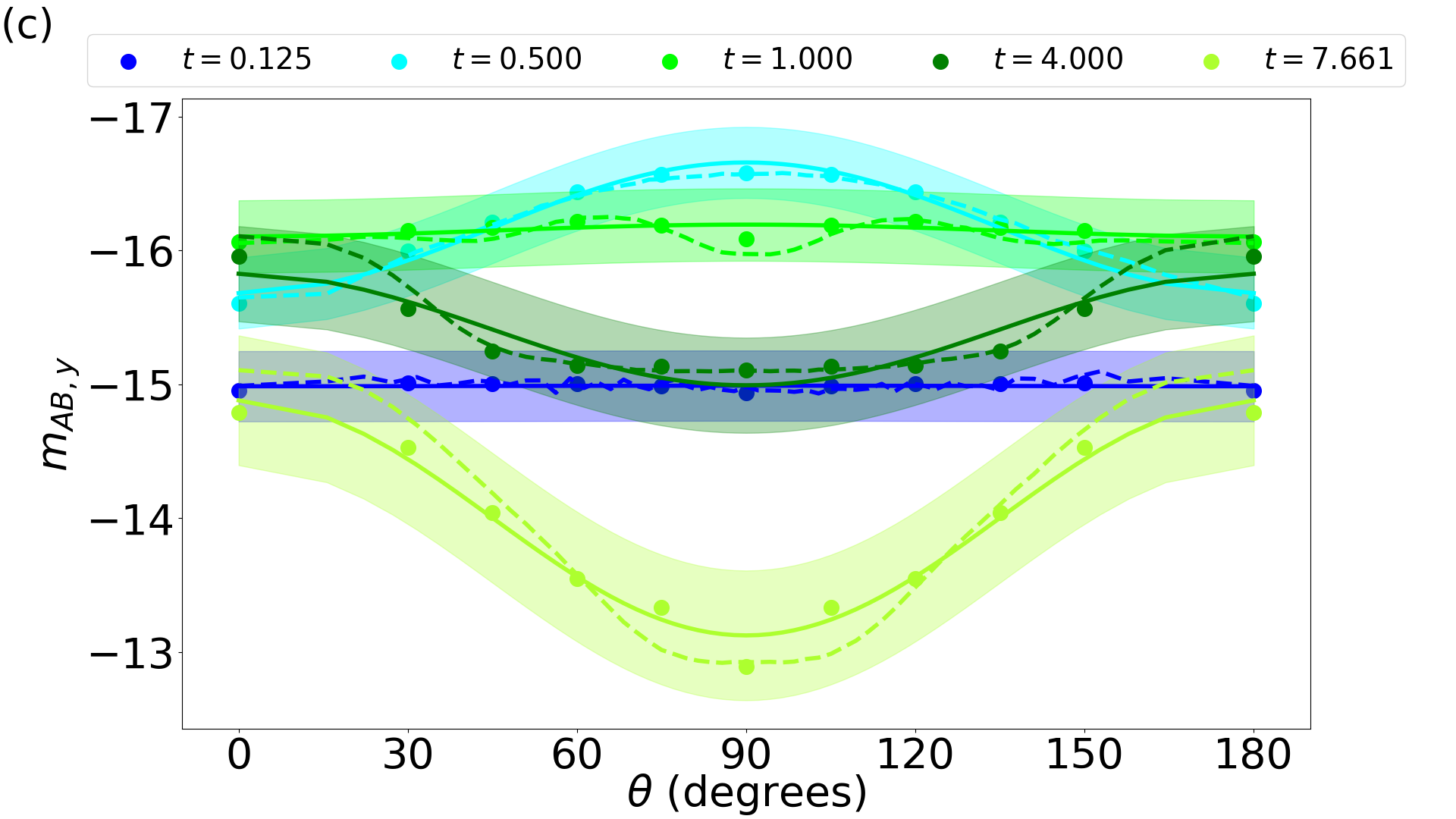}
\includegraphics[width=0.925\columnwidth]{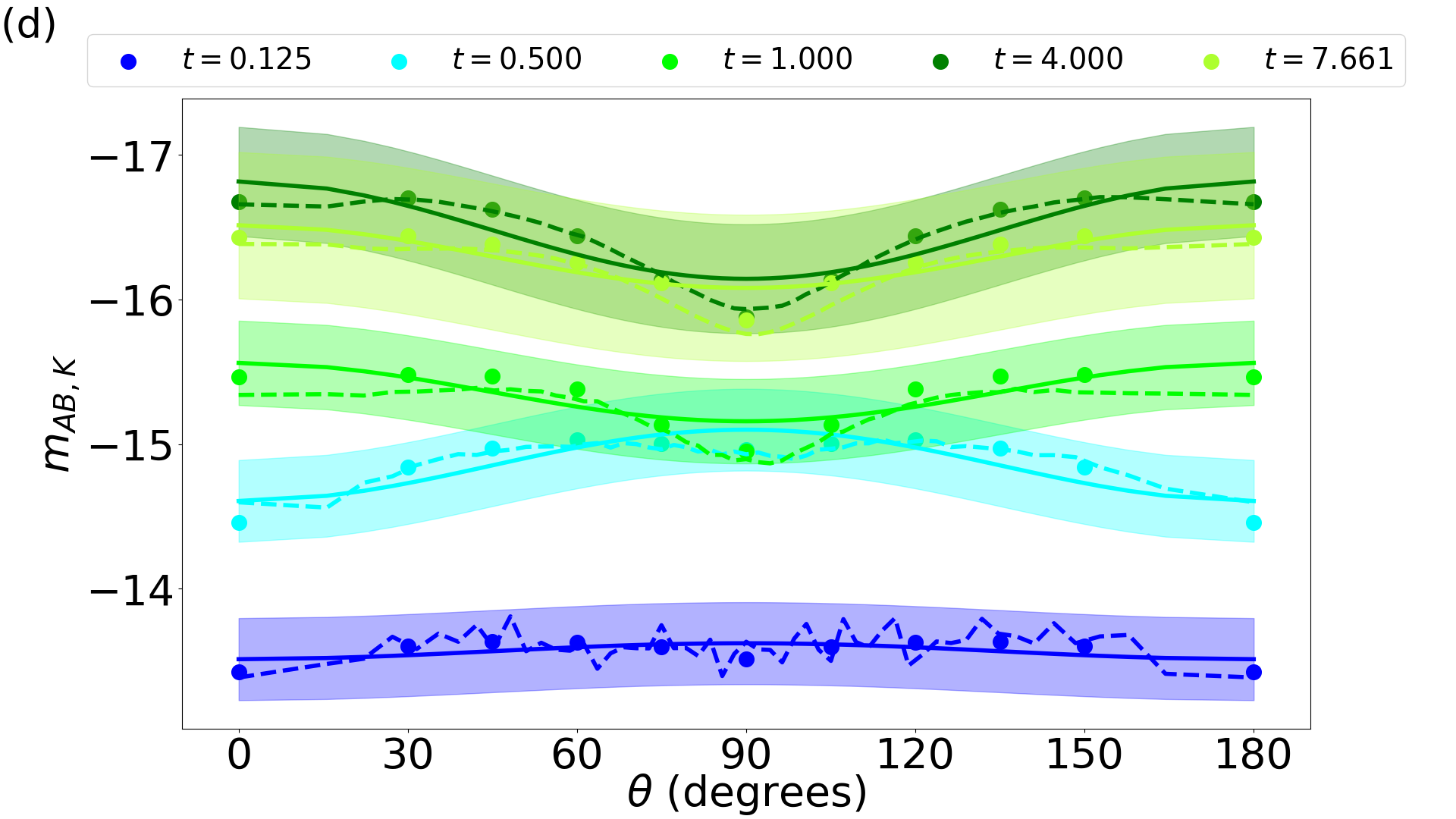}
\caption{\label{fig:AngleInterpolation} \textbf{Interpolation of $g$-, $y$-, and $K$-band luminosity at different viewing angles}: This figure
compares the $g$-, $y$-, and $K$-band luminosity at select times as a function of viewing angle. The solid points represent fixed angles at which the
different families of models were trained. The solid lines connecting the points indicate the interpolated prediction of the
angular variation at some given time in the light curve. The dashed lines represent the simulation data and show the true angular 
variation.
The shaded regions denote the $1\sigma$ error estimate derived from our Gaussian process fit versus time, extended in angle.
}
\end{figure}

All of the interpolated light curves discussed thus far have been trained at some fixed viewing angle. 
In Figure \ref{fig:AngleInterpolation}, we explore the interpolation of several families of models, each of which was 
trained using simulation data at a different viewing angle. The symmetry of the ejecta across the orbital plane allows for
the assumption that any angular variation between $0$ and $\pi/2$ can simply be mirrored across the symmetry axis. 

Figure \ref{fig:AngleInterpolation} indicates that the first day post-merger does not introduce much angular variation and, as such,
is quite well predicted even when interpolating across only 11
angles. After 1 day, the luminosity across different angles begins to change considerably as the peanut-shaped wind
ejecta becomes more dominant.
Particularly at late times, there is a strong angular variability which manifests near the orbital plane, most strongly apparent in
the blue ($g$) and near infrared ($K$) bands.  In the blue bands, the angular variation reflects lanthanide curtaining; in
the red bands, the angular variation reflects red emission from the late-peaking red dynamical ejecta. 
[At the latest times and faintest luminosities along the equatorial plane, numerical uncertainty in our Monte Carlo simulations are apparent in the
light-curve results.]
In all panels, the solid band denotes an estimated error bar from our GP fit in time, extended in angle.

\subsection{Predictive Accuracy versus time, angle grid sizes}

To better understand the systematic limitations and computational inefficiencies introduced by our stitched-time
interpolation grid, we investigated the accuracy of our fits when only using a subset of the time or angular grid.

First, we consider a simple analysis of loss of predictive accuracy as the number of GP interpolators used to make a surrogate light curve is decreased. We denote $t \in T$ as the subset of 
times represented by the GP interpolators used to make a prediction, $T$ as the total available number of time points, and thus interpolators, which can be used to make a light curve, and $\bar{t}$ 
as all the other times in $T$ which are not represented by $t$ such that $t \cap \bar{t} = 0$ and $t \cup \bar{t} = T$.

Thus, when using any number of interpolators at times $t \in T$ which is less than the total number of possible time points $T$, we first generate predictions $y(t)$ with the chosen subset of 
interpolators. These predictions $y(t)$, along with the times $t$ at which the predictions were made, are then used as inputs  for \texttt{SciPy}'s UnivariateSpline method 
from which the remainder of the light curve $z(t) = f({t}, y(t))$ is constructed, where in this last case the function can be evaluated $\forall t \in T$.

Figure \ref{fig:residuals_vs_nintps} shows how the average residual between on-sample light curve predictions and the respective simulation data changes
as a function of the number of time points used as the base for constructing the time-interpolated light curve.  For the
current scheme, we can remove up to roughly 75\% of the initial set of time points without substantially diminishing our
overall accuracy. Future work will explore smarter selection of representative time points in an effort to further reduce
the number of interpolators which can be removed without significant loss of accuracy.

\begin{figure}
\includegraphics[width=0.925\columnwidth]{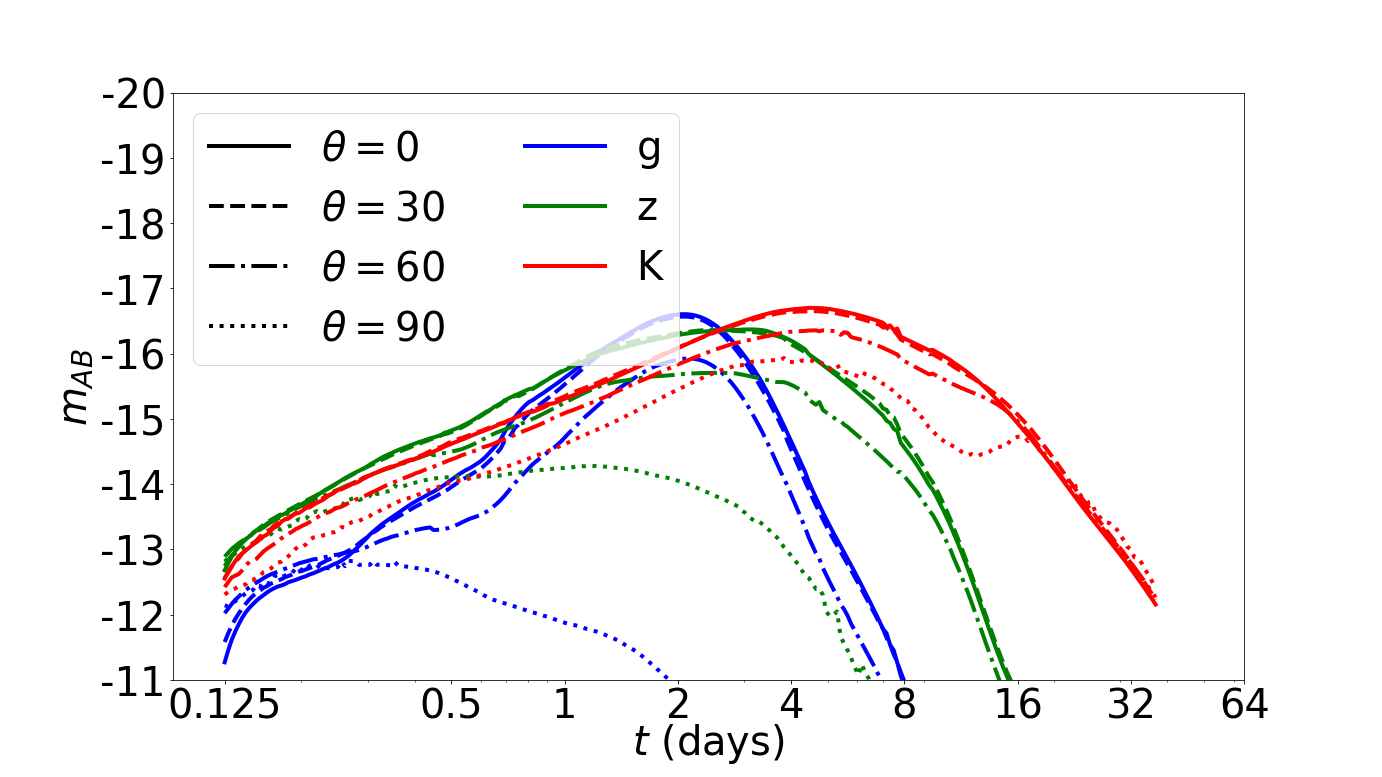}
\caption{\label{fig:AngularDependence:2} \textbf{Light curve versus time for selected angles and bands}: Comparison to Figure 7 of \cite{2020ApJ...889..171K}
indicating angular dependence of light-curve predictions across the $g$-, $z$- and $K$-bands.}
\end{figure}

\begin{figure}
\includegraphics[width=0.925\columnwidth]{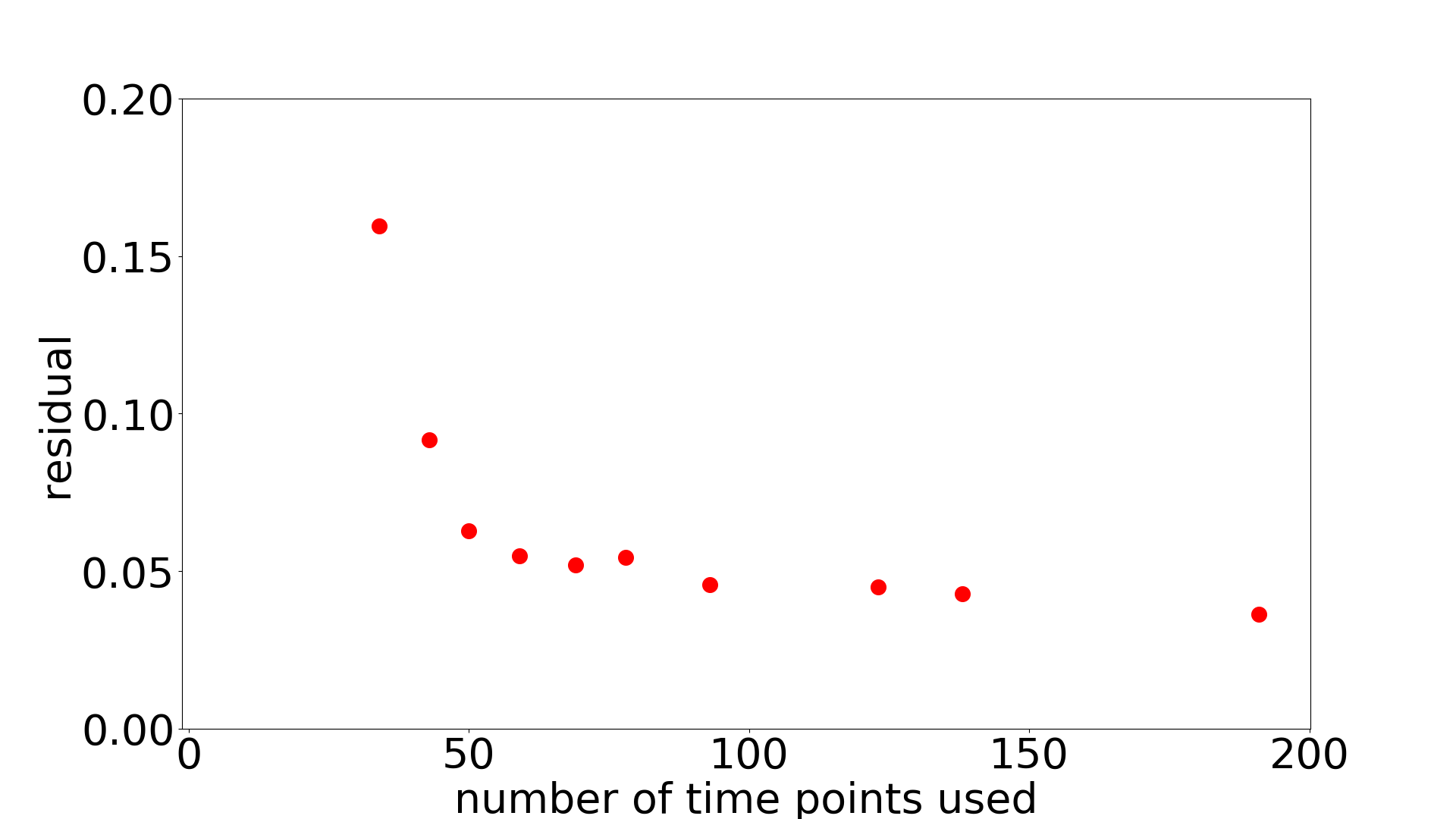}
\caption{\label{fig:residuals_vs_nintps} \textbf{Average residual as a function of number of considered time points}:
A plot of the average residuals between on-sample time-interpolated light curves and the respective simulation data as a function of how many time points are used to generate the light curves.
In each case, we drew the respective number of samples from a log-uniform distribution between the start and end time of our light curves.}
\end{figure}

\section{Parameter inference of radioactively-powered kilonovae }
\label{sec:PE}

In this section, we describe and demonstrate the algorithm we use to infer kilonova parameters given observations, using
the interpolated light-curve model above.   
Unless otherwise noted, for simplicity all
calculations in this section assume the kilonova event
time and distance are known parameters.  We likewise assume observational errors are understood and well
characterized by independent Gaussian magnitude errors in each observation, and that our model families include the
underlying properties of the source (i.e., we neglect systematic modeling errors due to the parameters held constant in
our simulation grid:  morphology, initial composition, et cetera).

\subsection{Framework and validation}
As in many previous applications of Bayesian inference  to infer parameters of kilonovae
\cite{gwastro-mergers-em-CoughlinGPKilonova-2020,2018MNRAS.480.3871C,2019MNRAS.489L..91C,2017ApJ...851L..21V,2017Natur.551...75S},
we seek to compare the  observed magnitudes  $x_i$ at evaluation points $i$ (denoting a combination of band and time) to a
continuous model that makes predictions $m(i|{\bm \theta})$ [henceforth denoted by $m_i({\bm \theta})$ for brevity] which depend on some model parameters $\theta$.  Bayes
theorem expresses the posterior probability $p({\bm\theta})$ in terms of a prior probability $p_{\rm
  prior}({\bm\theta})$ for the model parameters $\bm\theta$ and a likelihood ${\cal L}(\theta)$ of all observations,
given the model parameters, as 
\begin{equation}
p({\bm \theta}) = \frac{{\cal L}({\bm \theta}) p_{\rm prior}({\bm \theta})}{
  \int d {\bm \theta} {\cal L}({\bm \theta}) p_{\rm prior}({\bm \theta})
}
\end{equation}
Unless otherwise noted, for simplicity we assume the source sky location, distance, and merger time are known.
We adopt a uniform prior on the ejecta velocity $v/c\in[0.05,0.3]$ and a log uniform  prior on the ejecta masses
$m/M_\odot \in [10^{-3},0.1]$.  
We assume the observations have Gaussian-distributed \emph{magnitude} errors with presumed known observational
(statistical) uncertainties $\sigma_i$, convolved with
some additional unknown systematic uncertainty $\sigma$, so that 
\begin{equation}
    \ln \mathcal{L}(\bm{\theta}) = -0.5 \sum_{i=1}^n \left [ \frac {(x_i - m_i(\bm{\theta}))^2} {\sigma_i^2 + \sigma^2} + \ln(2 \pi (\sigma_i^2 + \sigma^2)) \right ]
\end{equation}
where the sum is taken over every data point in every band used in the analysis.   In tests, we  treat $\sigma$ as an uncertain model
parameter, de facto allowing  for additional systematic observational uncertainty (or for some systematic theoretical
uncertainty).  For our GP surrogate models, we set $\sigma$ to the estimated GP model error.
Unlike prior work, we eschew Markov-chain Monte Carlo, instead constructing the posterior distribution by direct Monte
Carlo integration as in  \cite{2015PhRvD..92b3002P,gwastro-PENR-RIFT}.  To efficiently capture correlations, we employ a
custom adaptive Monte Carlo integrator; see  \citet{gwastro-RIFT-Update} for implementation details.
In Appendix \ref{ap:validate_pe}, we describe several tests we performed to validate this inference technique using
synthetic kilonova data drawn from a previously published semianalytic kilonova model.    Our tests include recovering
the parameters of a hundred  synthetic kilonova sources.
In future work, we will demonstrate how our parameter inference method can be incorporated efficiently and
simultaneously with gravitational wave (GW)
parameter inference with the rapid iterative fitting (RIFT) parameter estimation pipeline \cite{gwastro-PENR-RIFT}.

\subsection{Inference with surrogate kilonova model}

\begin{figure}
\includegraphics[width=\columnwidth]{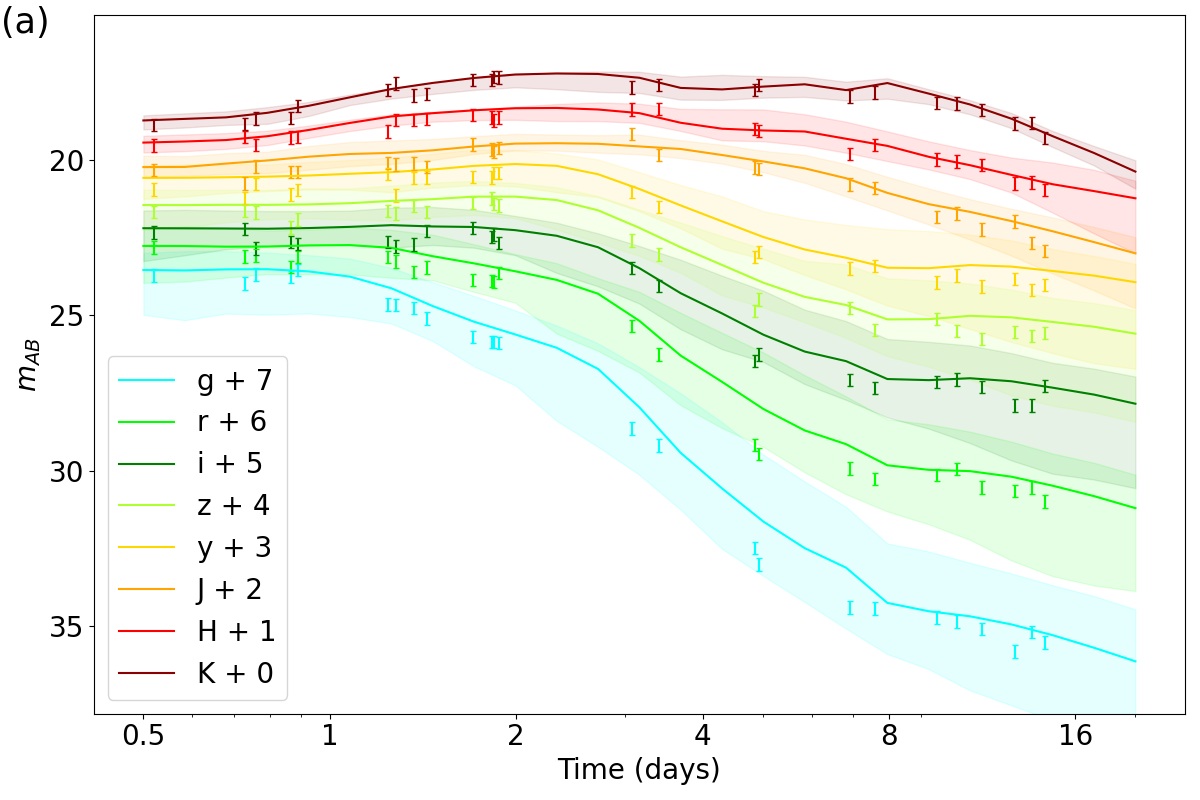}
\includegraphics[width=\columnwidth]{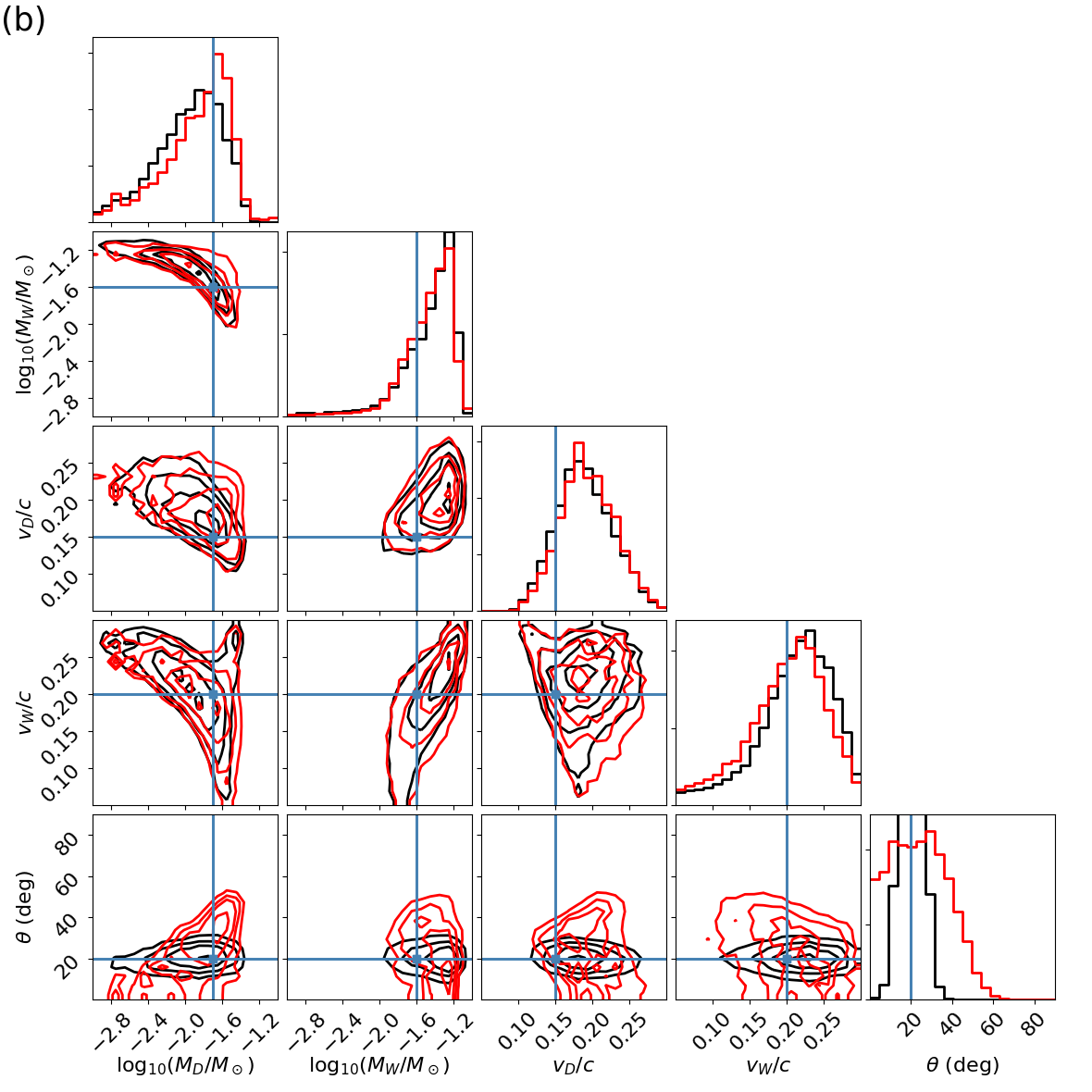}
\caption{\label{fig:pedemo:interp}\textbf{Synthetic source recovery with surrogate model: }
Recovery of a parameters of a known  two-component surrogate kilonova model, using inference based on our interpolated model.
Solid black curves show results adopting a strong angular prior motivated by radio observations of GW170817.
\emph{Top panel}: Synthetic light cuve data in several bands.
\emph{Bottom panel}: Inferred distribution of the four model parameters, and viewing angle. The blue cross denotes the
injected values.    Red contours show results without adopting a prior on observing angle; black contours show results
inferred when adopting a prior on viewing angle consistent with observations of GW170817.
}
\end{figure}

\begin{figure}
\includegraphics[width=\columnwidth]{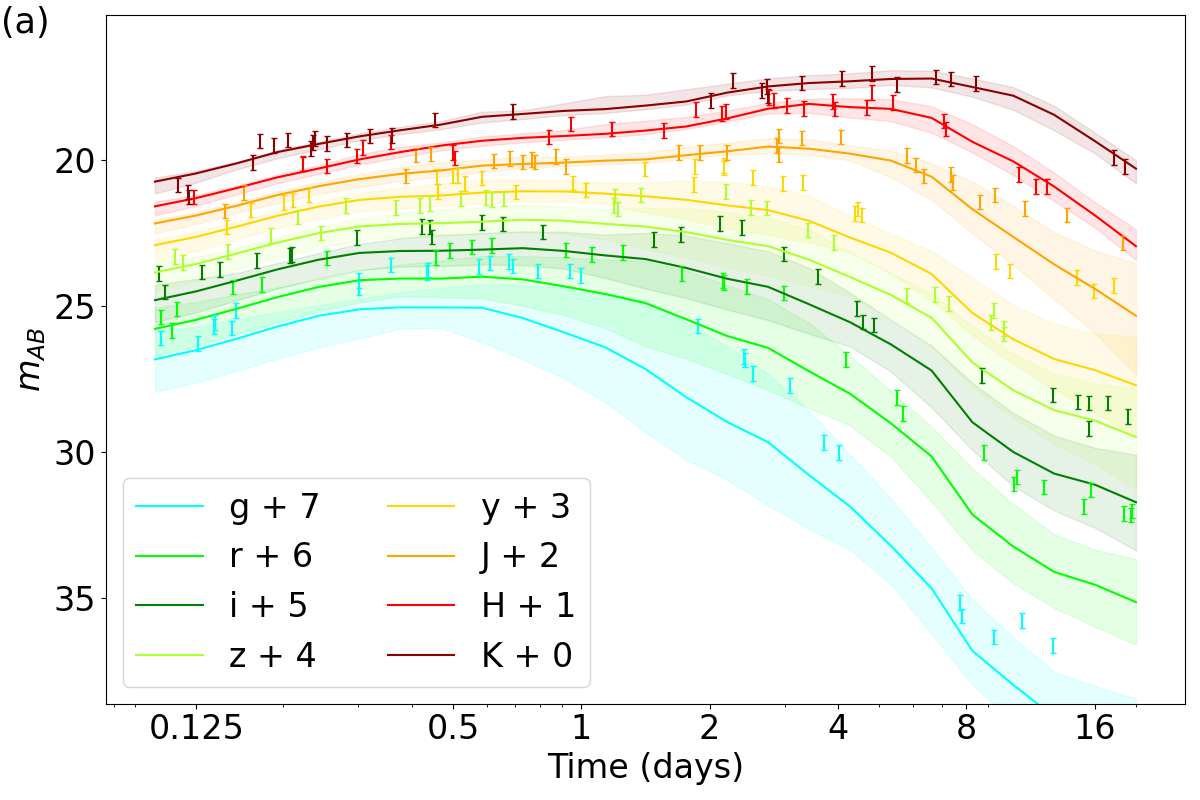}
\includegraphics[width=\columnwidth]{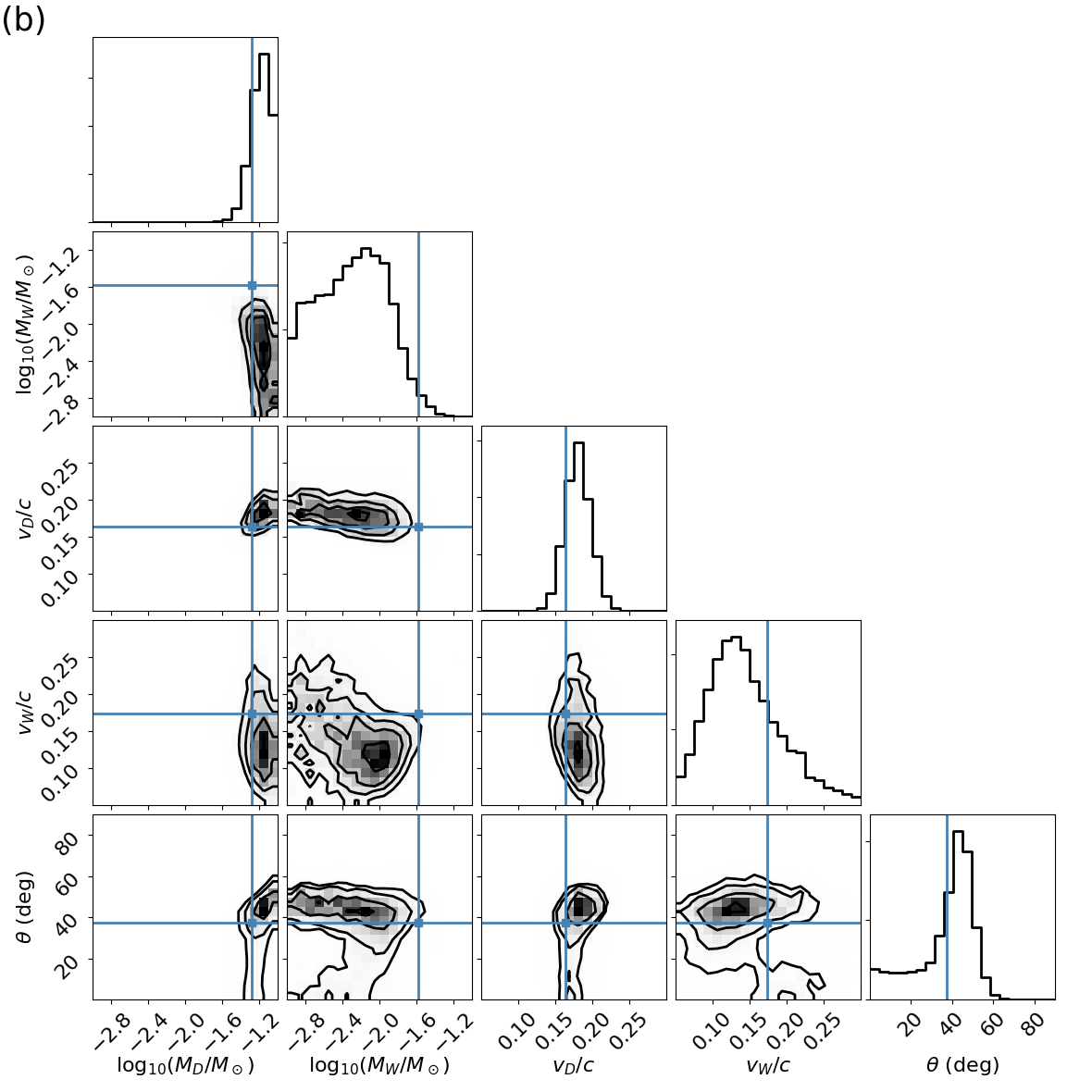}
\caption{\label{fig:pedemo:interp_on_sim}\textbf{Simulation parameter  recovery with surrogate model: }
Recovery of a parameters of a known  two-component kilonova \emph{simulation}, using inference based on our interpolated
model. The parameters corresponding to the relevant simulation are $M_{d}=0.052780 M_{\odot}$, $v_d=0.164316 c$, $M_{w}=0.026494 M_{\odot}$,
and $v_w=0.174017 c$.
\emph{Top panel}: Synthetic light curve data in several bands.
\emph{Bottom panel}: Inferred distribution of the four model parameters. 
}
\end{figure}

\begin{figure}
\includegraphics[width=\columnwidth]{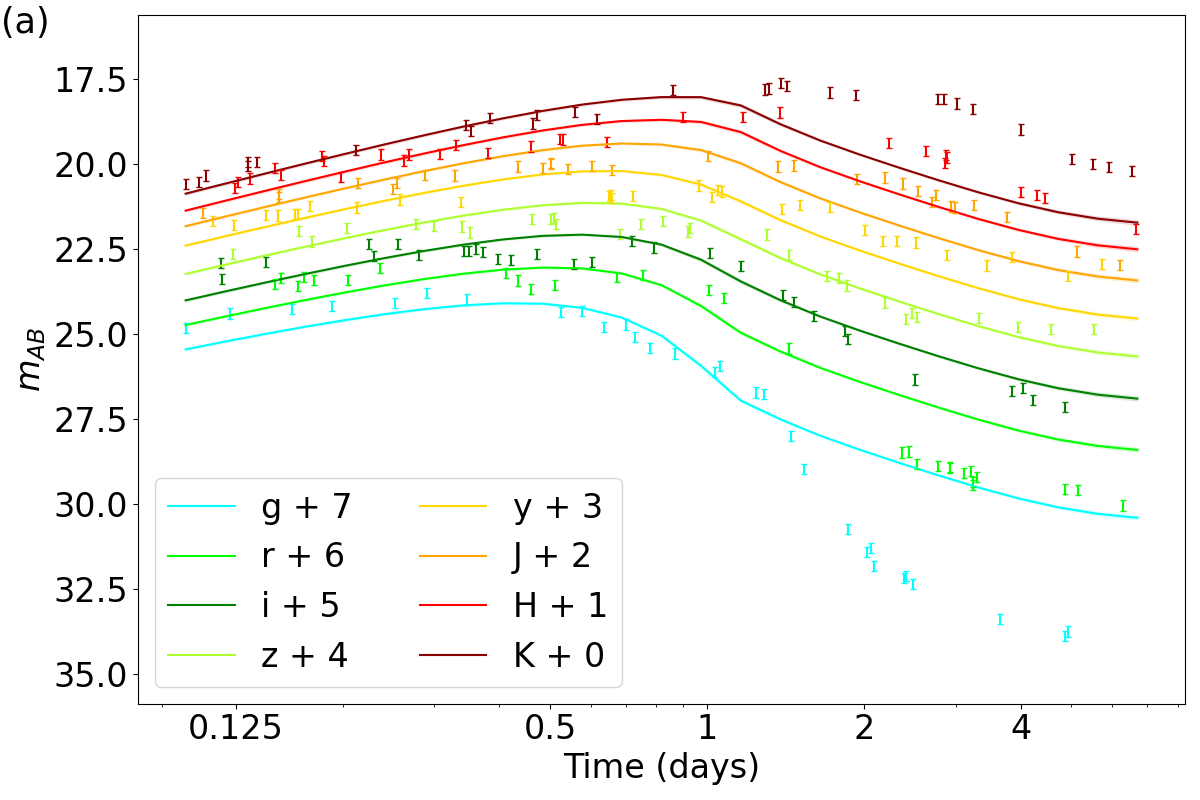}
\includegraphics[width=\columnwidth]{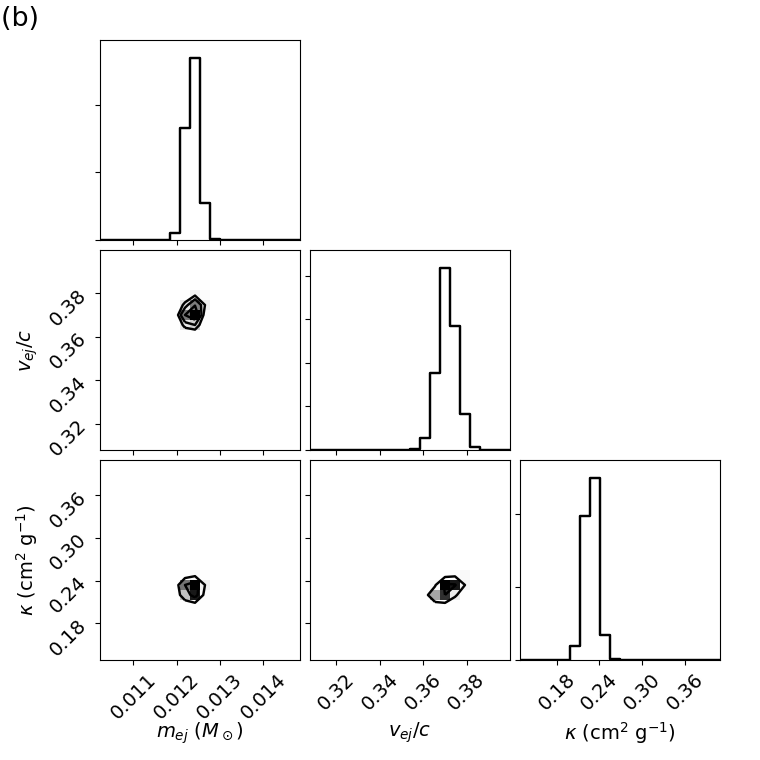}
\caption{\label{fig:pedemo:model_on_sim}\textbf{Simulation parameter  recovery with analytic model: }
Recovery of a parameters of a known  two-component kilonova \emph{simulation}, using inference based on the simplified
analytic model described in the appendix.  The analytic model cannot fit our simulation data.  While only a one-component
fit is shown, similar results arise when employing multiple components.
The parameters corresponding to the relevant simulation are $M_{d}=0.01 M_{\odot}$, $v_d=0.3 c$, $M_{w}=0.01 M_{\odot}$,
and $v_w=0.3 c$.
\emph{Top panel}: Synthetic light curve data in several bands, including error bars on both the synthetic data and
posterior light curve predictions.   The analytic model cannot fit our simulated data well.
\emph{Bottom panel}: Inferred distribution of the four model parameters. The blue cross denotes the injected
values.
}
\end{figure}

Figure \ref{fig:pedemo:interp} demonstrates parameter inference using our surrogate light curves, for a synthetic source
generated using our own model.  As expected, we can recover a known source, including constraining the viewing angle $\theta$.
Figure \ref{fig:pedemo:interp_on_sim} performs a similar test, but now using a specific simulation, without
  interpolation.  As expected given our adopted systematic error,  we recover the simulation parameters.
Finally, Figure \ref{fig:pedemo:model_on_sim} repeats the test above, using the semianalytic model described in the
appendix.  This comparison emphatically demonstrates large systematic differences between this semianalytic model and
our detailed simulations.  

\subsection{Example: GW170817}

\begin{figure*}
\includegraphics[width=\textwidth]{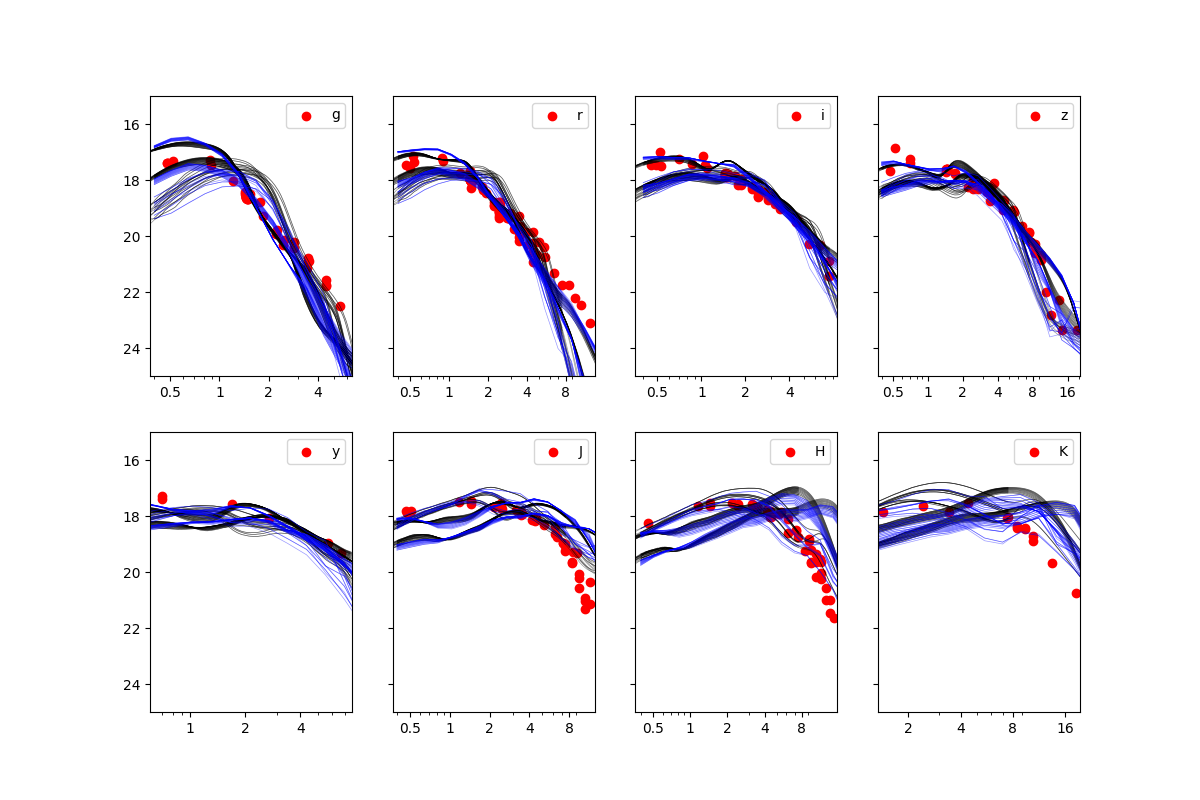}
\caption{\label{fig:170817:SimulationsOnly}\textbf{Comparison to GW170817: Simulations only}: This figure shows the
  results of direct comparison between all our  simulations and our observations.  Each simulation and
  angle is assigned a likelihood; the top 50 highest likelihood simulations  are shown (black), compared with
  the observational data (red).   The 50 black curves are drawn from a total number of potential candidate angles and simulations of $54\times 448$.  For comparison, the blue curves show our continuously-interpolated model, evaluated at
  the same parameters as the underlying simulations.   We emphasize the identified simulations are at the edges of our
  simulation domain, where interpolation error is most severe.
}
\end{figure*}

SuperNu-based kilonova models have already been successfully used to interpret GW170817, 
though as noted previously these models have  a  rapid  falloff  in  the  late-time optical magnitudes that is not present in
the observations;   see  \cite{tanvir17}.
 Because
of the close proximity of GW170817, only distance modulus (but not redshift)  corrections are needed to translate our
predictions to apparent magnitudes which can be directly compared to electromagnetic observations.
Observational results are taken from \citep{2017ApJ...851L..21V}'s compilation of
  photometry reported in 
\cite{2017Natur.551...64A,tanvir17,2017Natur.551...71T,2017ApJ...848L..17C,2017Sci...358.1570D,2017Natur.551...75S,2017ApJ...848L..24V,2017Natur.551...67P,2017Sci...358.1559K,2017Sci...358.1574S,2017PASJ...69..101U}.
Figure \ref{fig:170817:SimulationsOnly} shows the results of directly comparing our extended simulation archive directly
to observations of GW170817, selecting for simulations (parameters and angles) with the highest overall likelihood.  The
solid black curves in these figures show the 50 highest-likelihood configurations, where the likelihood requires
simultaneously reproducing all observed bands.  Except for reddest three bands
(JHK), many simulations compare extremely favorably to the observations. 
The parameters of these simulations, however, do not  represent the optimal parameters of this model family: because our placement
algorithm minimizes interpolation error, the selected points preferentially occur at the edges of our domain. 
Finally, for the reddest band ($K$), our fits exhibit notable systematic uncertainty relative to the underlying
simulation grid.

\begin{figure*}
\includegraphics[width=\columnwidth]{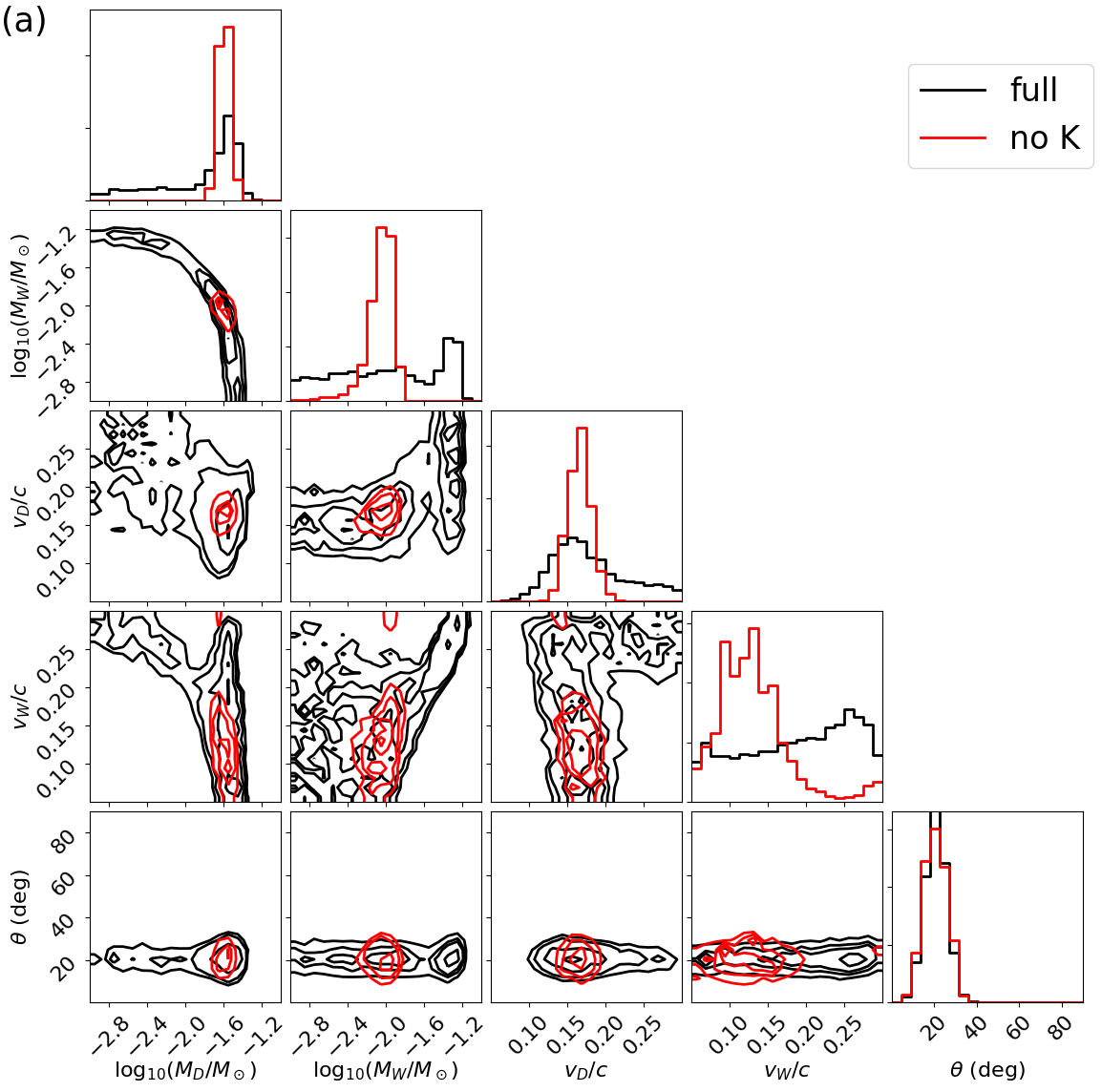}
\includegraphics[width=\columnwidth]{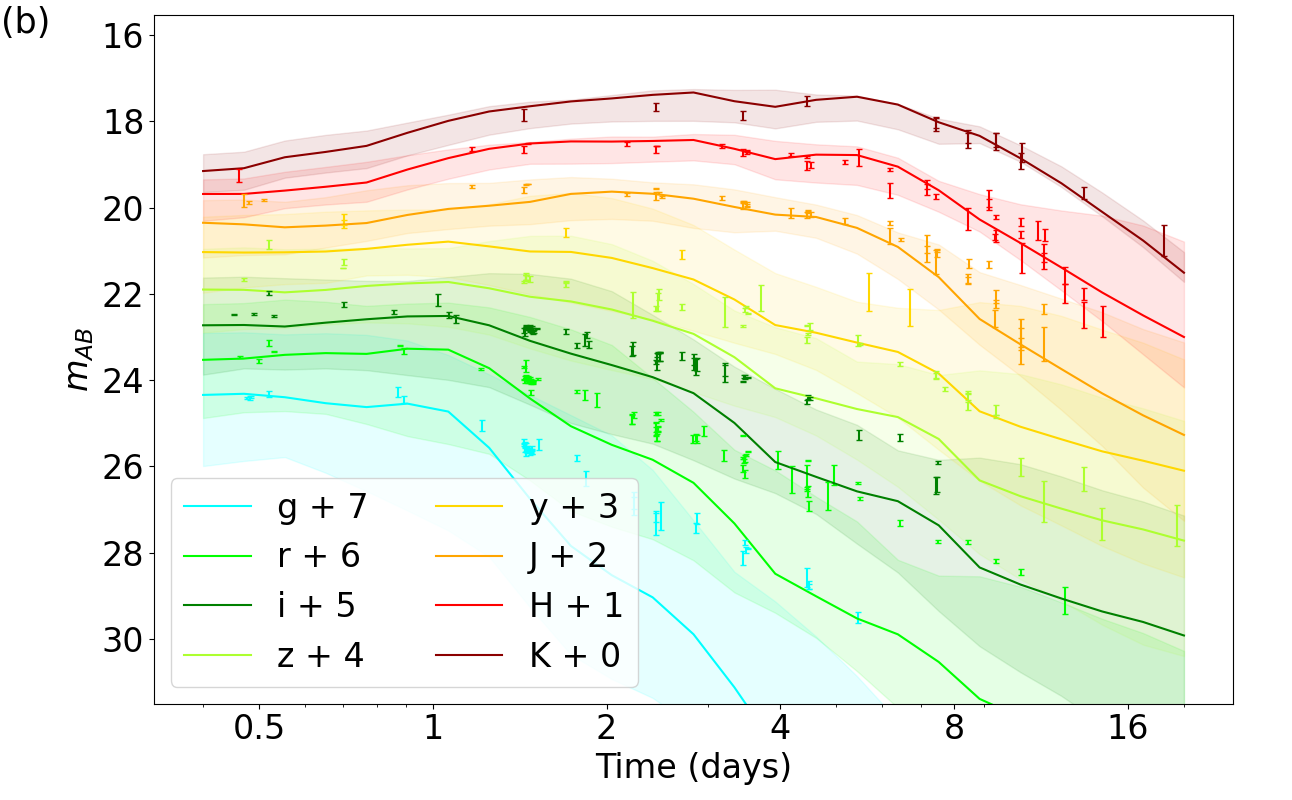}
\caption{\label{fig:170817:ToyModel}\textbf{Comparison to GW170817}:  The left panel shows the results of interpreting observations of GW170817
  using  our surrogate light curve model and adopting a strong angular prior $\theta\simeq 20\unit{deg}$.  In the left
  panel, the solid black shows inferences using all observing data, while the red curve omits $K$-band observations.
The right panel shows inferred light curves corresponding to  the full data set analysis (i.e., the light curves
correspond to the black contours in the left panel).
}
\end{figure*}

We have performed parametric inference on GW170817 using our surrogate light curve model to the underlying SuperNu
results.  
Motivated by the direct comparisons above, we perform two analyses.  In the first, we use all observing data at all
times.
In the second, we omit the reddest ($K$) band.
Figure \ref{fig:170817:ToyModel} shows the results of these comparisons.  Because of the systematic fitting
uncertainties at late times, we highlight the analysis omitting $K$-band observations as our preferred result.
Though previously-reported inferences about ejecta masses cover a considerable dynamic range  (see, e.g., Fig. 1 in \cite{2019EPJA...55..203S}),  our inferred masses are
qualitatively consistent with selected previous estimates including  previous inferences with similar SuperNu models 
\cite{tanvir17} and recent surrogate models adapted to simplified multidimensional radiative transfer
\cite{2020Sci...370.1450D}.
Notably, however, we infer a large amount of ``dynamical'' (red, lanthanide-rich)  ejecta mass  (i.e.,
$M_{ej}\simeq O(1/30) M_\odot$), more dynamical ejecta than wind, and the velocities for the dynamical and wind component are inverted relative to customary expectations (i.e., $v_d<v_w$). 
Our dynamical and wind component masses differ from typical semi-analytical treatments (see e.g. \cite{Nicholl21}).

We also weakly constrain the misalignment angle between the outflow and the line of sight to be consistent with
independent late-time radio observations, which (weakly supplemented with gravitational wave constraints)
constrain the opening angle between the jet and the line of sight to be roughly $20$
degrees 
\cite{2018Natur.561..355M,2019NatAs...3..940H, Evans_2017, 2017Natur.551...71T, tanvir17, Troja_2020}; see also  \cite{2020Sci...370.1450D} for previous, weaker constraints on binary alignment from kilonova observations.
Reanalyzing the light curves using this prior information, imposed as a Gaussian prior
on $ \theta$ with mean $20 \unit{deg}$ with $\sigma_\theta=5\unit{deg}$, we find modestly improved overall constraints on the
ejecta; see also also, e.g., Figure 2 of \cite{2020Sci...370.1450D}  for previous, weaker constraints derived using joint radio and
kilonova observations.

\section{Discussion}
\label{sec:discussion}

We have demonstrated that our  surrogate models can be operationally compared to real kilonova observations, allowing us
to deduce what the range of parameters for the original simulation family best fits the observations.  In this section,
we emphasize several systematic limitations of our approach, to more clearly distinguish the ways in which the answers
so obtained could differ from a description of physical reality.   When possible, we comment on ways in which these
systematic limitations could be mitigated with future work.

First and foremost, our surrogate models introduce some modest bias, being an imperfect representation of the
simulations they mimic.   We have demonstrated that these errors are relatively small (see Figures \ref{fig:off_sample_interp}
and \ref{fig:AngleInterpolation}).
In this work, we principally employed two standard interpolation methods (GP and RF interpolation) to construct
synthetic light curves.   Recent substantial advances in machine learning have led to many new algorithms and
architectures for adaptive learning and interpolation.  Our prior work suggests that neural networks can also usefully
interpolate kilonova light curves \cite{RisticThesis}, which we will describe at greater length in future work.
Other groups have also successfully produced surrogate light curves with modest error.
Previously, surrogate light curves have been produced by interpolating the coefficients $c_g(\Lambda)$ of a basis-function
expansion $\log L_\alpha(t|\Lambda) = \sum_g c_g(\Lambda)\phi_g(t)$
\cite{gwastro-mergers-em-CoughlinGPKilonova-2020,2018MNRAS.480.3871C}, with the appropriate basis functions identified
by principal component analysis of the raw simulation output.  
Because of the prohibitive cost of GP on large data sets, this analysis had to decimate input data to enable interpolation.
Our reference-time method offers several notable advantages.  The most important advantage is that our method is embarrassingly parallel--interpolations
 at every reference time can be performed independently, without need to select suitable basis functions
in advance--and completely decoupled between time samples.  Our interpolation is also inherently local in time, so
artifacts inherited from late-time simulation data of low-photon-count light curves cannot contaminate  our estimates of
early-time behavior.  Finally, our method can in principle be applied to all available data, without decimation,
particularly when we employ other interpolation techniques.

Second, we adopt simulations with an imperfect model of the relevant opacities and nuclear physics.  For example,  we
have adopted a conventional nuclear mass and decay model to predict nuclear heating and element 
abundances \cite{Korobkin_2012, 2018MNRAS.478.3298W}. We also find that although our detailed multifrequency opacities yield a more
realistic representation of the physics in the system, the assumption of thermalization breaks down much sooner than
anticipated. 
Uncertainties in nuclear physics can play a substantial role in kilonova light curves
\cite{2020arXiv201011182B,2021ApJ...906...94Z}.  Given sufficient simulations, surrogate light curves can be constructed for a wide range of nuclear
inputs.   We defer a systematic treatment of nuclear physics uncertainties and non-LTE opacities to later work. 

Third, we employ simulations with phenomenological initial conditions, that are not initialized with the appropriate
orientation-dependent distribution of mass, velocity, and composition versus time.   More suitable initial conditions could be
provided by   detailed disk simulations \cite{2019PhRvD.100b3008M,2020arXiv201107176D}.

Fourth, we do not (and cannot) initialize our simulations with initial data that is set by physics of the merger.
Unlike previous work \cite{2019MNRAS.489L..91C,2018MNRAS.480.3871C}, we have not adopted a relationship between our two-component ejecta
parameters and the   progenitor masses $m_1,m_2$, motivated by  substantial uncertainty
in the nuclear equation of state and remnant lifetime \cite{2015PhRvD..91f4059S,2017PhRvD..96l1501D,LIGO-GW170817-EOSrank,2020JHEAp..27...33L}. 
Even the best-available fits have considerable systematic uncertainty \cite{2020PhRvD.101j3002K}.
Similarly, we have not adopted assumptions about the lifetime of any hypermassive remnant and the duration of neutrino
illumination \cite{2020JCAP...04..045D}, nor have we incorporated radiation from any associated jet  \cite{2021MNRAS.500.1772N}.
Instead, given substantial systematic uncertainty in merger simulations (relative to the small amount of ejecta), we treat the ejecta purely phenomenologically, implicitly allowing
for  many potential nuisance parameters to characterize the outflow.

Fourth, our models were trained on a subset of simulations with fixed ejecta morphologies and mass fractions for both components. 
The predictive capability of our interpolations is restricted to two-component models represented by the parameters in Table \ref{tbl:grid_params}.
Our parameter estimation inherits these limitations and should thus also be considered for bias which stems from the selected subset of models in
our interpolation training library.

Finally, we note that GW170817 was likely an exceptional case which contributed sufficient quantities of observational data for extremely informative parameter inference. Our methodology is still applicable in cases with sparser light curve data; however, the level of detail in the inference results will vary based on the aforementioned sparsity.

\section{Conclusions}
\label{sec:conclude}

We have adaptively constructed detailed anisotropic models for kilonovae that cover a four-dimensional
space describing two components' masses and velocities.  From these models, we have constructed  surrogate multiband
light curves which can be evaluated continuously over this space.  We have demonstrated how our model can be used for
kilonova source parameter inference, including the kilonova associated with GW170817.  
All of our input data products, fitted light curves and the code we used to produce them are available at
\texttt{https://github.com/markoris/surrogate\char`_kne}.  The underlying full simulations are available at
\texttt{https://ccsweb.lanl.gov/astro/transient/ \
transients\char`_astro.html}.

Though we limited our study to a specific set of assumptions, this analysis is an important stepping stone towards a
better understanding of kilonova systematics.  Recently, several studies have demonstrated that several physical
assumptions can notably impact the deduced light curve.   However, these impacts could have effects that are partially
degenerate with modest shifts in ejecta properties.  To understand the practical impact of these uncertainties, in
future work we will   employ our parameterized models with these sources of error. 
In this work we emphasized inference on only phenomenological kilonova parameters.  Several studies have demonstrated
the value in using multimessenger information to more tightly constrain parameters like source inclination
(see, e.g.,  \cite{2018Natur.561..355M,2019NatAs...3..940H,2020NatCo..11.4129C,2020Sci...370.1450D}),
even without adopting strong assumptions about the relationship between ejecta and progenitor masses.  
With such assumptions even stronger constraints have been widely explored.
In future work we will show how the electromagnetic inference strategy applied here can be tightly and efficiently integrated with the
RIFT parameter inference engine, enabling concordance inference about multimessenger sources. 

\begin{acknowledgments}
ROS and MR acknowledge support from NSF AST 1909534.
EAC acknowledges financial support from the IDEAS Fellowship, a research traineeship program funded by the National Science Foundation under grant DGE-1450006.
CF, CF,  AH, OK, and RW were supported by the US Department of Energy through the Los Alamos National Laboratory.
Los Alamos National Laboratory is operated by Triad National Security, LLC, for the
National Nuclear Security Administration of U.S.\ Department of Energy (Contract No.\ 89233218CNA000001).
Research presented in this article was supported by the Laboratory Directed Research and Development
program of Los Alamos National Laboratory under project number 20190021DR.
This research used resources provided by the Los Alamos National Laboratory Institutional Computing
Program, which is supported by the U.S. Department of Energy National Nuclear Security Administration
under Contract No.\ 89233218CNA000001.
This document has been assigned LA-UR-21-24289.
\end{acknowledgments}
\clearpage\appendix

\section{Validation of parameter inference method}
\label{ap:validate_pe}
\subsection{Simple analytic kilonova model}

To validate our parameter inference codes, we implemented a standard semianalytic kilonova model previously presented in  \cite{2017ApJ...851L..21V}.
This model consists of  single-component, two-component, and three-component models, combined by flux addition and not
allowing for  anisotropy. 
In the following equations, $M$ is the $r$-process ejecta mass (in $M_\odot$)  and $v$ is the ejecta velocity.
Note that for now we assume the ejecta consists entirely of $r$-process material, so $M$ is the full ejecta mass.
The radioactive heating rate at time $t$ is given by \cite{Korobkin_2012}:
\begin{equation}
    L_\text{in}(t) = 4 \times 10^{18} M \times \left [ 0.5 - \pi^{-1} \arctan \left ( \frac {t - t_0} {\sigma} \right ) \right ]^{1.3} \text{ erg s}^{-1}
\end{equation}
where $t_0 = 1.3$ s and $\sigma = 0.11$ s are constants.

Only a fraction of $L_\text{in}$ powers the kilonova, given by the thermalization efficiency $\epsilon_\text{th}$.
This is approximated analytically in \cite{2016ApJ...829..110B}:
\begin{equation}
    \epsilon_\text{th}(t) = 0.36 \left [ e^{-a t} + \frac {\ln(1 + 2 b t^d)} {2 b t^d} \right ]
\end{equation}
The parameters $a$, $b$, and $d$ are constants that depend on the ejecta mass and velocity; an interpolation of Table 1 in \cite{2016ApJ...829..110B} is used in the model.
The bolometric luminosity is calculated via \cite{Chatzopoulos_2012}
\begin{equation}
    L_\text{bol}(t) = \frac {2} {t_d} \exp{\left ( \frac {-t^2} {t_d^2} \right ) }
                    \int_0^t L_\text{in} \epsilon_\text{th} \exp{\left ( \frac {t^2} {t_d^2} \right ) } \frac {t} {t_d} dt
\end{equation}
where $t_d$ is the diffusion timescale, $t_d = \sqrt{2 \kappa M / \beta v c}$, $\kappa$ is the opacity, and $\beta = 13.7$ is a dimensionless constant related to the ejecta's geometry.

Light curves are calculated by assuming the kilonova behaves as a blackbody photosphere that expands at a velocity $v$.
The blackbody temperature is generally defined by its bolometric luminosity; however, once it cools to a critical temperature $T_c$, the photosphere recedes into the ejecta and the temperature remains fixed.
The photosphere temperature is
\begin{equation} \label{T_c}
    T_\text{phot}(t) = \max \left [ \left ( \frac {L_\text{bol}(t)} {4 \pi \sigma_\text{SB} v^2 t^2} \right )^{1/4}, T_c \right]
\end{equation}
where $\sigma_{SB}$ is the Stefan-Boltzmann constant.
When $T_\text{phot} > T_c$, the photosphere radius is simply $R_\text{phot} = v t$.
When $T_\text{phot} = T_c$ (i.e. the photosphere has receded into the ejecta), the photosphere radius is
\begin{equation}
    R_\text{phot}(t) = \left ( \frac {L_\text{bol}(t)} {4 \pi \sigma_\text{SB} T_c^4} \right )^{1/2}
\end{equation}

The flux density at frequency $\nu$ is given in \cite{2019LRR....23....1M}:
\begin{equation}
    F_\nu(t) = \frac {2 \pi h \nu^3} {c^2} \frac {1} {\exp{(h \nu / k T_\text{photo}(t))} - 1} \frac {R_\text{photo}^2(t)} {D^2}
\end{equation}
where $D$ is the source distance.
We use a fixed fiducial distance of $D = 10$ pc to calculate $F_\nu(t)$, then calculate AB magnitude with a distance modulus if necessary.

To compute multi-component light curves we assume each component has a photosphere that evolves independently of the others.
The total flux density is the sum of the flux densities of the individual components.
The version of the model implemented in the code uses three components with fixed opacities (a blue component with $\kappa = 0.5$ cm$^2$ g$^{-1}$, a purple component with $\kappa = 3$, cm$^2$ g$^{-1}$ and a red component with $\kappa = 10$ cm$^2$ g$^{-1}$).

\subsection{Validation test: recovery of synthetic kilonova}
\begin{figure}[!ht]
    \centering
    \includegraphics[width=\columnwidth]{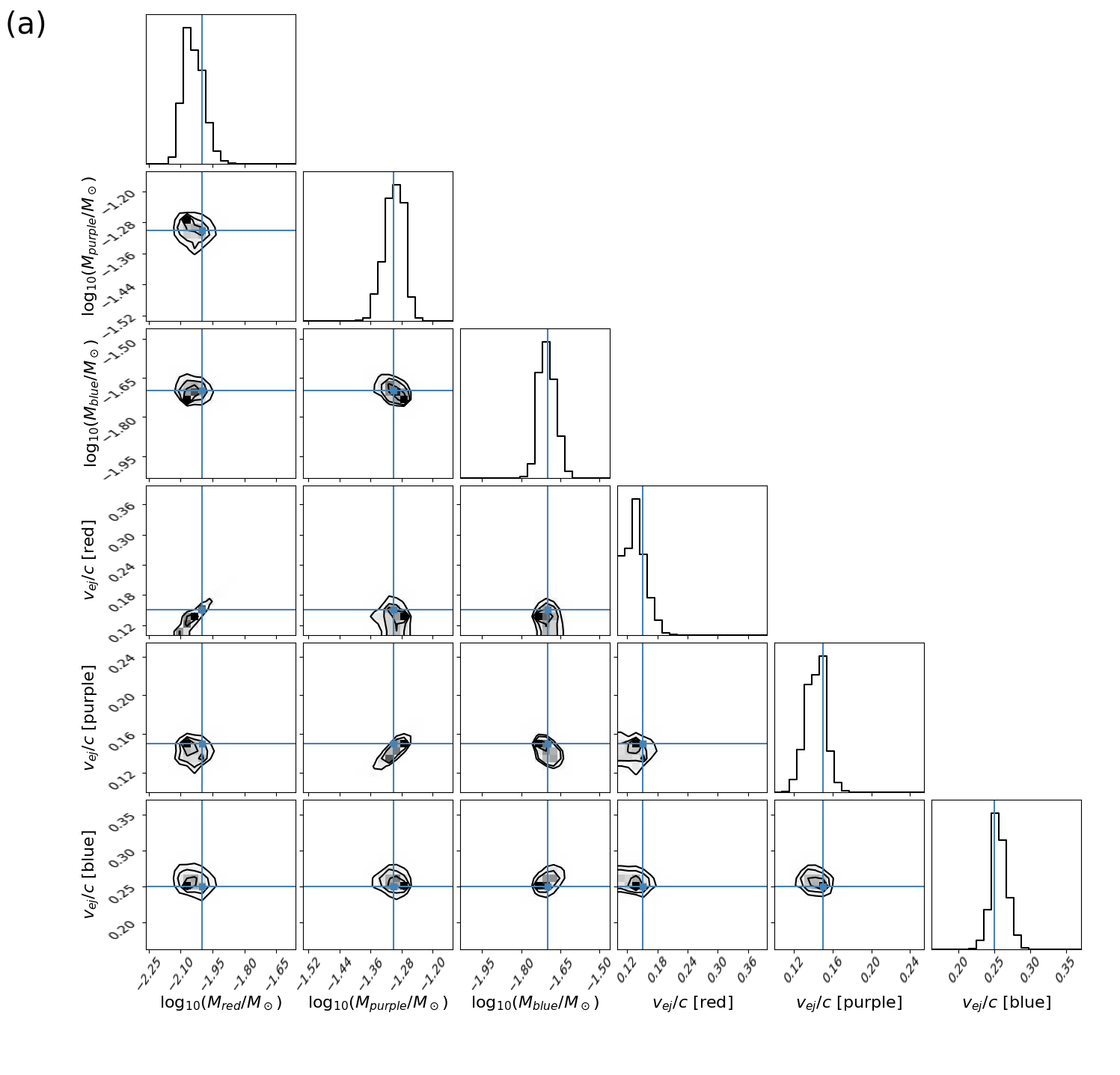}
    \includegraphics[width=\columnwidth]{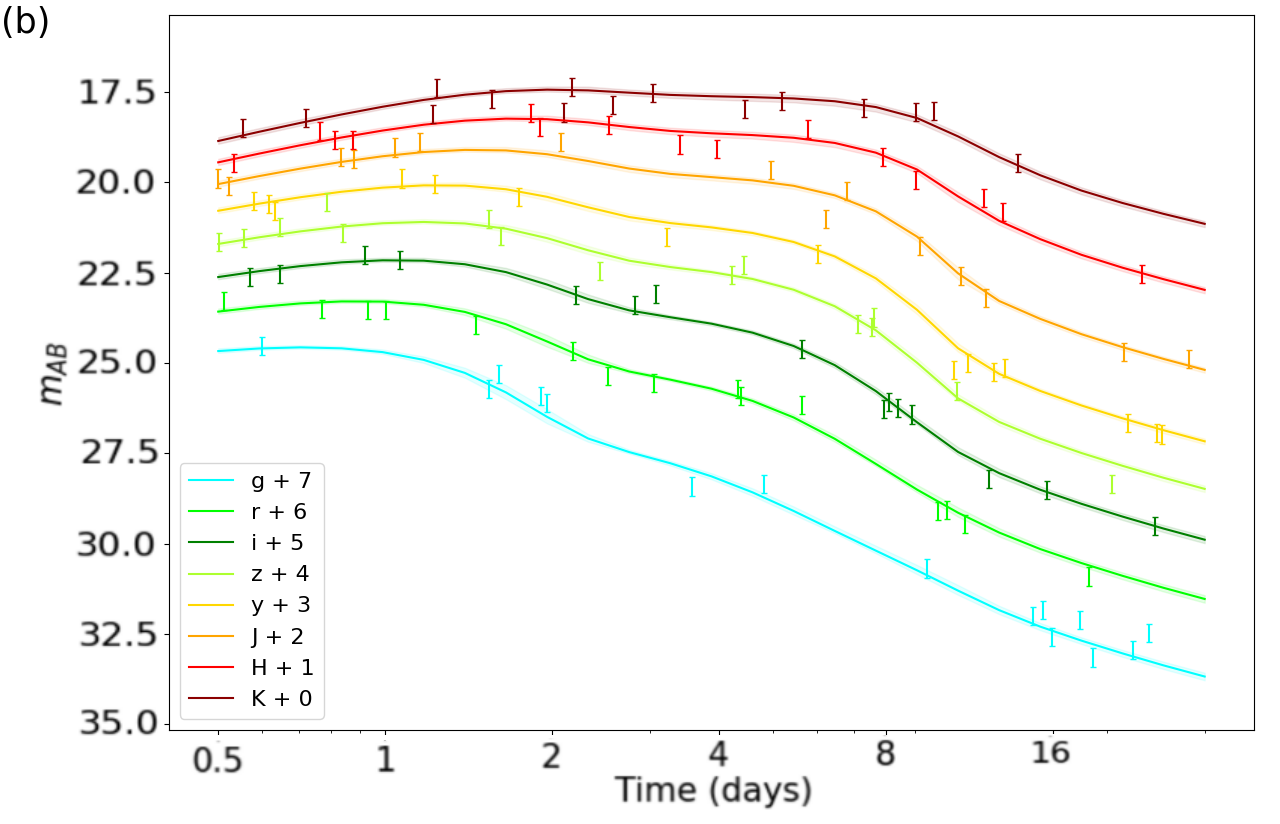}
    \caption{\textbf{Analytic kilonova model injection: }\emph{Top panel}:Posterior distribution for the injection/recovery test (the blue lines show the true
      parameter values). For this example, $T_c$ was fixed for each component and the distance was fixed to 40.0 Mpc.
\emph{Bottom panel}: Fake photometry data and light curve models for the injection/recovery test. The solid line shows
the light curve model evaluated at the maximum-likelihood parameters.  Both the data and light curves include a confidence interval.}
    \label{fig:validate_single}
\end{figure}

Figure \ref{fig:validate_single}  illustrates our implementation of this synthetic model and our parameter inference
code.   The error bars on the bottom panel show a densely-sampled synthetic multiband light curve with plausible
kilonova parameters for all three components.   The top panel shows a standard corner plot representing the parameters
of our synthetic kilonova with thin blue lines; one-dimensional marginal posterior distributions on the diagonal for
each parameter; and  two-dimensional marginal posterior distributions in the bottom corner.

\subsection{Validation test: random synthetic kilonova and probability-probability plot}

\begin{figure}
\includegraphics[width=\columnwidth]{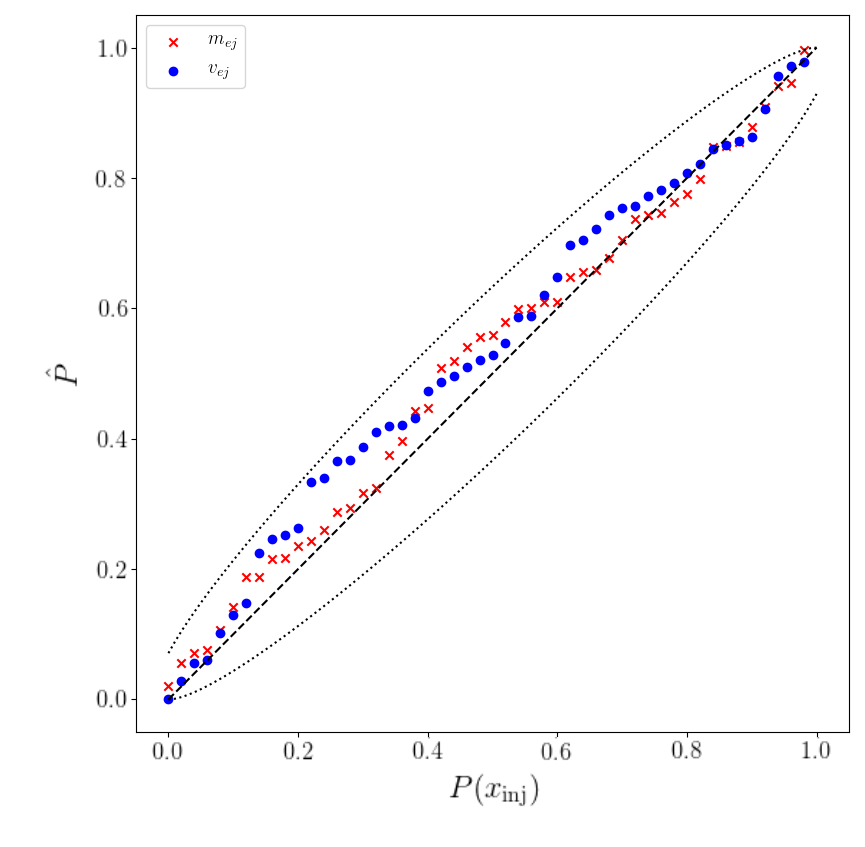}
\caption{\label{fig:validate_pp}\textbf{Probability-probability (P-P) tests: }
 P-P plot for 100 synthetic injections generated with one random component.
}
\end{figure}

We also demonstrated our inference technique using 100 randomly generated light curves, drawn uniformly from the same
priors we use for inference and incorporating noise consistent with our Gaussian noise model.  Using these 100 synthetic events and inferences, we can construct a
probability-probability (P-P) plot \cite{mm-stats-PP}, which  corroborates that the one-dimensional marginal distributions are consistent.
Figure \ref{fig:validate_pp} shows the results of our analysis.  %
More extensive tests of the underlying integration algorithm, including PP plots using more complex and higher
dimensional models, are reported elsewhere \cite{gwastro-RIFT-Update}.

\clearpage

\bibliography{bibliography,gw-astronomy-mergers-ns-gw170817,LIGO-publications}

\end{document}